\documentclass[%
 reprint,
superscriptaddress,
 amsmath,amssymb,
 aps
]{revtex4-2}

\usepackage{graphicx}
\usepackage{dcolumn}
\usepackage{bm}
\usepackage{tabularx}
\usepackage{amsmath}
\usepackage{cancel}

\begin{document}

\title{Evidence of a ``current-mediated'' turbulent regime in space and astrophysical plasmas}

\author{Luca Franci}
\email{l.franci@qmul.ac.uk}
\affiliation{School of Physics and Astronomy, Queen Mary University of London, London, UK.}
\affiliation{INAF, Osservatorio Astrofisico di Arcetri, Firenze, Italy.}
\author{Daniele Del Sarto}
\affiliation{Institut Jean Lamour UMR 7198 CNRS, Universit\'e de Lorraine, Nancy, France}
\author{Emanuele Papini}
\affiliation{Dipartimento di Fisica e Astronomia, Universit\`a degli Studi di Firenze, Sesto Fiorentino, Italy}
\affiliation{INAF, Osservatorio Astrofisico di Arcetri, Firenze, Italy.}
\author{Alice Giroul}
\affiliation{School of Physics and Astronomy, Queen Mary University of London, London, UK.} 
\author{Julia E. Stawarz}
\affiliation{Department of Physics, Imperial College London, London, UK}
\author{David Burgess}
\affiliation{School of Physics and Astronomy, Queen Mary University of London, London, UK.}
\author{Petr Hellinger}
\affiliation{Astronomical Institute,The Czech Academy of Sciences, Prague, Czech Republic.}
\author{Simone Landi}
\affiliation{Dipartimento di Fisica e Astronomia, Universit\`a degli Studi di Firenze, Sesto Fiorentino, Italy}
\affiliation{INAF, Osservatorio Astrofisico di Arcetri, Firenze, Italy.}
\author{Stuart D. Bale}
\affiliation{Physics Department and Space Sciences Laboratory, University of California, Berkeley, CA. USA.}

\date{\today}

\begin{abstract}
How the turbulent energy cascade develops below the magnetohydrodynamic scales in space and astrophysical plasmas is a major open question. 
Here, we measure the power spectrum of magnetic fluctuations in Parker Solar Probe's observations close to the Sun and in state-of-the-art numerical simulations of plasma turbulence. Both reveal a power-law behavior with a slope compatible with $-11/3$ at scales smaller than the ion characteristic scales, steeper than what is typically observed in the solar wind and in the Earth's magnetosheath. We explain such behavior by developing a simple two-fluid model which does not require any kinetic processes nor electron-inertia effects. This is characterized by a significant contribution of the ion kinetic energy to the total turbulent energy cascade at sub-ion scales, although the dynamics is driven by the magnetic field through the current density. We expect that this regime may be relevant for a broad class of low-beta plasmas, e.g. the solar corona, non-relativistic magnetized jets and disks, and laboratory plasmas.
\end{abstract}

\maketitle



The solar wind, the collisionless plasma stream of charged particles that continuously flows outward from the Sun, is heated from thousands to millions of degrees over a distance of just a solar radius and accelerated up to speeds of 800 km/s.
The mechanisms responsible for such a behavior are not yet well understood.
In-situ spacecraft measurements close to the Earth suggest that plasma turbulence is a key ingredient for the solar wind dynamics and therefore a possible candidate for heating and accelerating its particles~\citep[e.g.,][]{Tu_Marsch_1995,Matthaeus_Velli_2011, Alexandrova_al_2013,Bruno_Carbone_2013,Sahraoui_al_2009}. The observed power spectra of the plasma and electromagnetic fluctuations exhibit a power-law behavior spanning many decades in frequency~\citep[e.g.,][]{Kiyani_al_2015,Podesta_al_2007}, thus providing evidence that an extended turbulent cascade is at play. This is responsible for transferring energy via nonlinear coupling from large down to smaller and smaller scales, through those characteristic of ions and electrons, where it is eventually dissipated by kinetic heating mechanisms other than particle collisions.

Particle heating and energy dissipation in the absence of collisions are also important phenomena in a number of different plasma environments over an extremely large range of scales, both inside the heliosphere (e.g., in the solar corona~\citep{Marsch2004}) and outside (e.g., in accretion disks~\citep{Balbus_Hawley_1998}, in the interstellar medium~\citep{Ferriere_2019}, and in galaxy clusters ~\citep{Schekochihin_Cowley_2006}). The solar wind is, however, unique among such systems, since it can be directly probed by means of spacecraft missions.

The NASA Parker Solar Probe (PSP) spacecraft mission~\citep{Fox_al_2016} has been approaching the Sun in an attempt to shed light on the origin of the solar wind. During its first five orbits, PSP has measured the plasma and electromagnetic fluctuations closer than any previous mission. PSP observations have already led to important discoveries, including dramatic magnetic field reversals on time scales of seconds or minutes~\citep{Bale_al_2019,Dudok_de_Wit_2020}, in-situ generated electromagnetic waves at ion resonant scales~\citep{Bowen_al_2020b,Huang_al_2020}, a significant increase in the energy transfer rate at magnetohydrodynamic scales with respect to that observed at 1 au~\citep{Bandyopadhyay_al_2020b}, an enhanced proton heating associated with large-amplitude turbulent fluctuations~\citep{Martinovic_al_2020}, and a steepening of the magnetic field spectrum below the magnetohydrodynamic scales~\citep{Bowen_al_2020c}.

Here we present the first evidence, both observational and numerical, for a plasma regime that occurs at scales just below the ion inertial length and is characterized by a power spectrum of the magnetic fluctuations with a slope (spectral index) compatible with $-11/3$. 
We also provide a theoretical interpretation based on a simple non-relativistic two-fluid model, which does not require any ion and electron kinetic processes (e.g., wave-particle
resonances) or any electron-inertia effects.
The model predictions are in agreement with recent spacecraft observations
and are supported by results of numerical simulations. The predicted regime occurs when the ion kinetic energy is not negligible with respect to the magnetic energy at sub-ion scales and provides an additional contribution to the energy cascade. The dynamics, however, is still guided by the evolution of the magnetic field via the current density. In this sense, we define this as a ``current-mediated'' turbulent plasma regime.
Based on our simulations, we expect that this is more likely to occur when the turbulent fluctuations are large and the plasma beta, i.e., the ratio of the plasma to the magnetic pressure, is low. As such conditions are more typically met in the near-Sun environment rather than further away, this might explain why it has not been identified before, while it seems to be frequent in PSP observations.

The current-mediated plasma regime can account for an enhanced sub-ion-scale energy transfer rate, due to the additional contribution of the ion kinetic energy, and possibly for a stronger turbulent heating close to the Sun than that predicted by alternative turbulent cascade models. This could shed light on the solar wind origin and, more in general, on particle heating and acceleration in collisionless space and astrophysical plasmas.

\section{Results}
\label{sec:results}

\subsubsection{Observational magnetic field spectra}
\label{sec:obs_spectra}
\begin{figure*}
\includegraphics[width=0.48\textwidth]{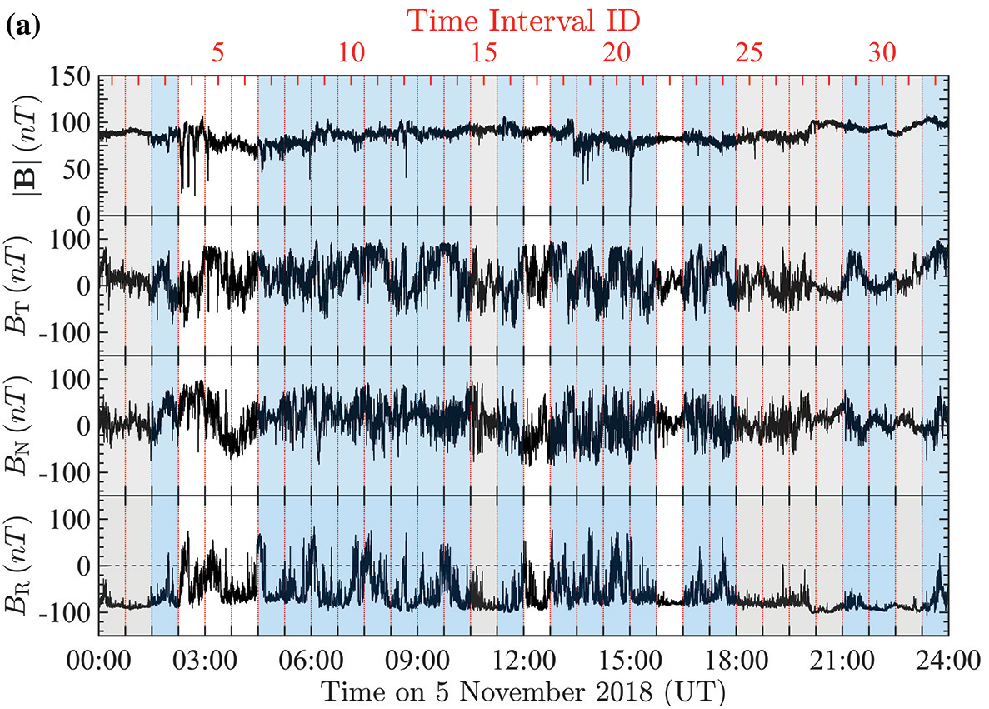}
\includegraphics[width=0.48\textwidth]{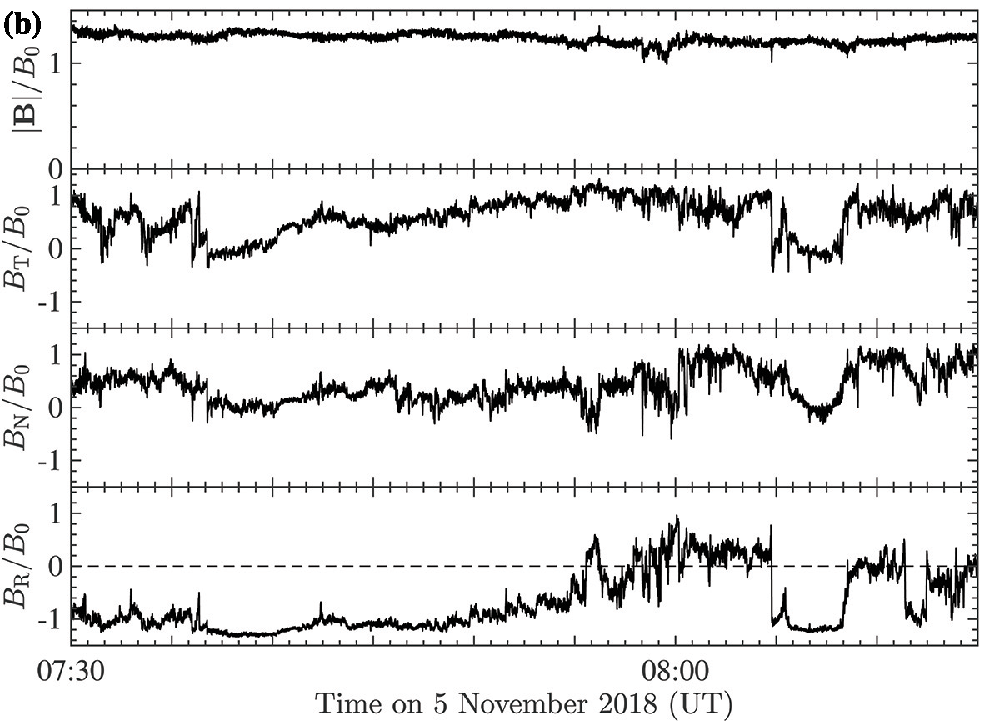}\\
\vspace{0.2cm}
\includegraphics[width=0.48\textwidth]{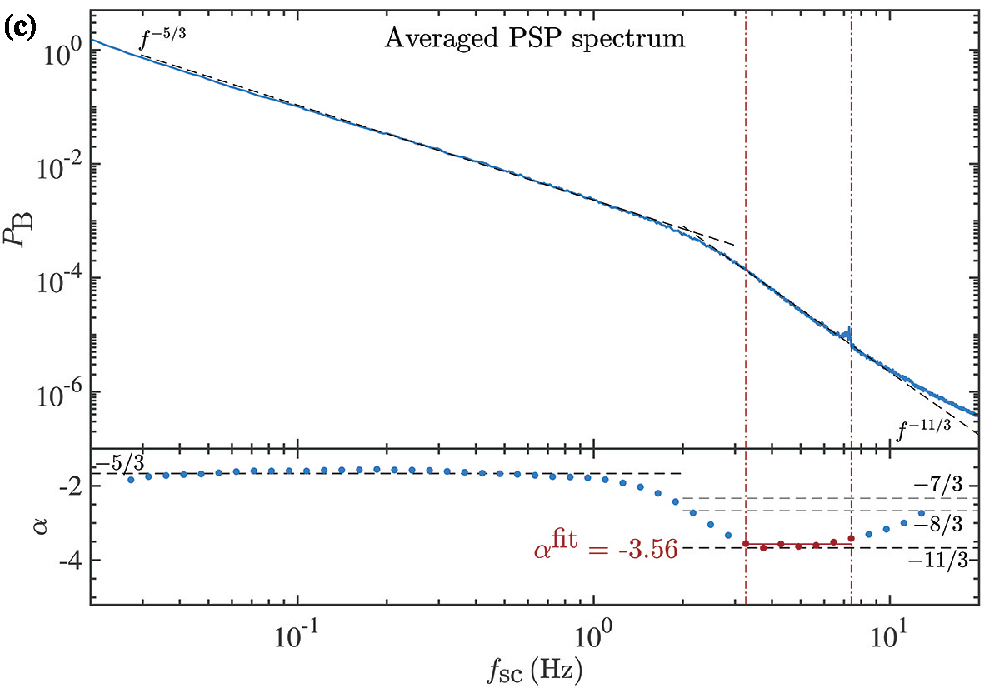}
\includegraphics[width=0.48\textwidth]{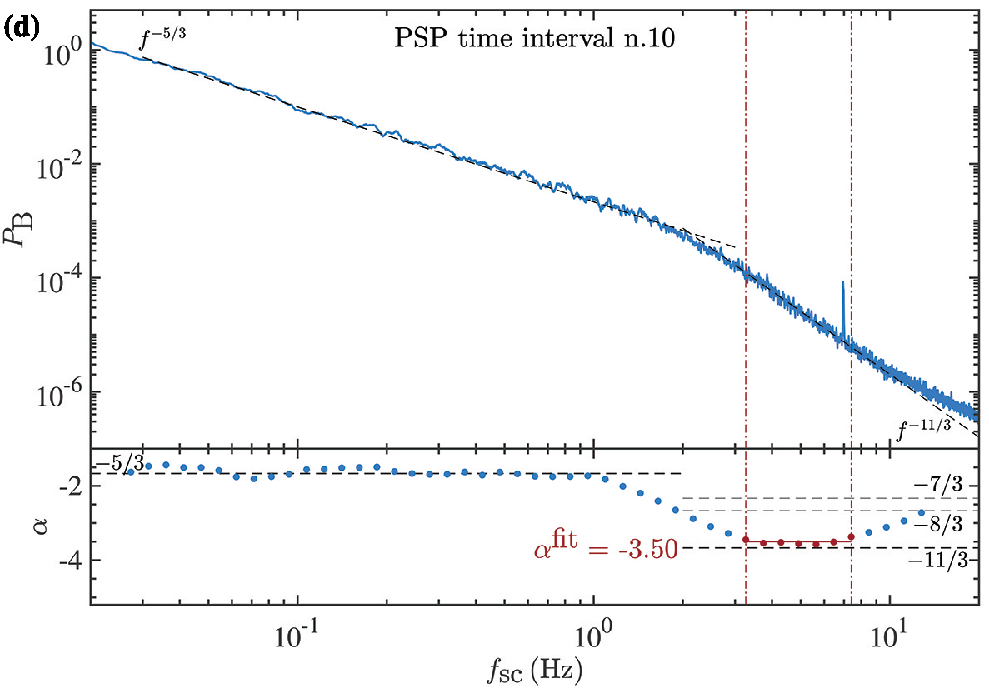}
\caption{Observational dataset: PSP observations of the magnetic field and its spectral behavior at about 36 solar radii from the Sun during its first perihelion on November 5, 2018 (00:00-24:00 UT). (\textbf{a}) 
Time series measurements of, from top to bottom, the magnetic field magnitude $|B|$ and its tangential ($B_\mathrm{T}$), normal ($B_\mathrm{N}$), and radial ($B_\mathrm{R}$) components. The day has been divided into 32 45-min intervals, limited by vertical red dotted lines and labeled with an ID on the top axis. (\textbf{b}) The same as in (\textbf{a}) but limited to the time interval \#10, (07:30-8:15 UT). All quantities here are normalized to the amplitude of the mean (vector) magnetic field $\bm{B}_0$ computed over the individual 45-min interval. (\textbf{c}) Top panel: Trace power spectrum of the magnetic field obtained by averaging all the 32 45-minute individual spectra, each normalized to its mean magnetic field. As a reference, power laws are superimposed with slope $-5/3$, classically predicted for the inertial-range turbulent cascade at supra-ion scales, and with $-11/3$, predicted by our model at sub-ion scales. Bottom panel: Local spectral index $\alpha$. Horizontal black dashed lines mark the classical asymptotic predictions for the turbulent cascade in the MHD inertial range, $-5/3$, and at sub-ion scales for a regular cascade, $-7/3$~\citep{Schekochihin_al_2009}, for an intermittent one, $-8/3$~\citep{Boldyrev_Perez_2012}, and our prediction, $-11/3$. The horizontal red line represents the fit of the local slopes over the range of frequencies limited by the two dot-dashed vertical red lines. (\textbf{d}) The same as in (\textbf{c}), but for time interval \#10, (07:30-8:15 UT). The spectrum has been normalized to the ambient magnetic field $B_0$ in the time interval.}
\label{fig:PSP}
\end{figure*}

The observational dataset consists of
high-resolution magnetic field measurements by the FIELDS fluxgate magnetometer~\citep{Bale_al_2016} on-board PSP~\citep{Fox_al_2016} at 36 solar radii from the Sun during its first perihelion on November 5, 2018 (00:00-24:00 UT) (see Sec.~\ref{subsec:methPSP} for further details). \\
The time series of the magnetic field magnitude and of its tangential, normal, and radial components for the full day are shown in Fig.~\ref{fig:PSP}a, from top to bottom. We choose to divide the dataset into 32 consecutive 45-minute intervals, as we intend to focus our attention on the spectral behavior at frequencies larger than $\sim 10^{-2}$ Hz. The spectral properties at larger frequencies for the first two PSP orbits, including the first perihelion, have been recently analysed and discussed in~\citet{Chen_al_2020}.
The first four of our intervals correspond to the ones discussed in Fig. 2 of ~\citep{Bale_al_2019}. Most, although not all, intervals contain reversals of the radial magnetic field component, corresponding to jets (sometimes also called ``switchbacks'')~\citep{Bale_al_2019,Dudok_de_Wit_2020}. Some intervals show instead a more quiet radial-field wind, with very small fluctuations of the radial magnetic field component and a quite constant magnitude. Such  behavior is evident in Fig.~\ref{fig:PSP}b, which show the same quantities as in Fig.~\ref{fig:PSP}a, limited to the time interval \#10 (07:30-8:15 UT).

The trace power spectrum of the magnetic field, $P_\mathrm{B}$, for interval \#10 is shown in the top panel of Fig.~\ref{fig:PSP}d (see Sec.~\ref{subsec:methFIT} for the definition of power spectrum). Its slope (spectral index) is well approximated by $-5/3$ above the ion characteristic scales (``supra-ion'' scales), over at least one full decade in frequency, and by $-11/3$ below (sub ion-scales). The latter behavior holds up to $\sim 10$ Hz, where the measurements from the fluxgate magnetometer get close to its noise floor.
In the bottom panel of the same figure, the local spectral index, $\alpha$, obtained by performing power-law fits over sliding windows of half a decade, is shown. 
The averaged spectrum, i.e., the result of averaging all the 32 spectra, each normalized to the mean field in the corresponding interval, further confirms this behavior (Fig.~\ref{fig:PSP}c): the local spectral index between $\sim3$ Hz and $\sim8$ Hz is $-3.56$.

\subsubsection{Numerical magnetic field spectra}
\label{sec:num_spectra}

\begin{figure*}
\includegraphics[width=0.5\textwidth]{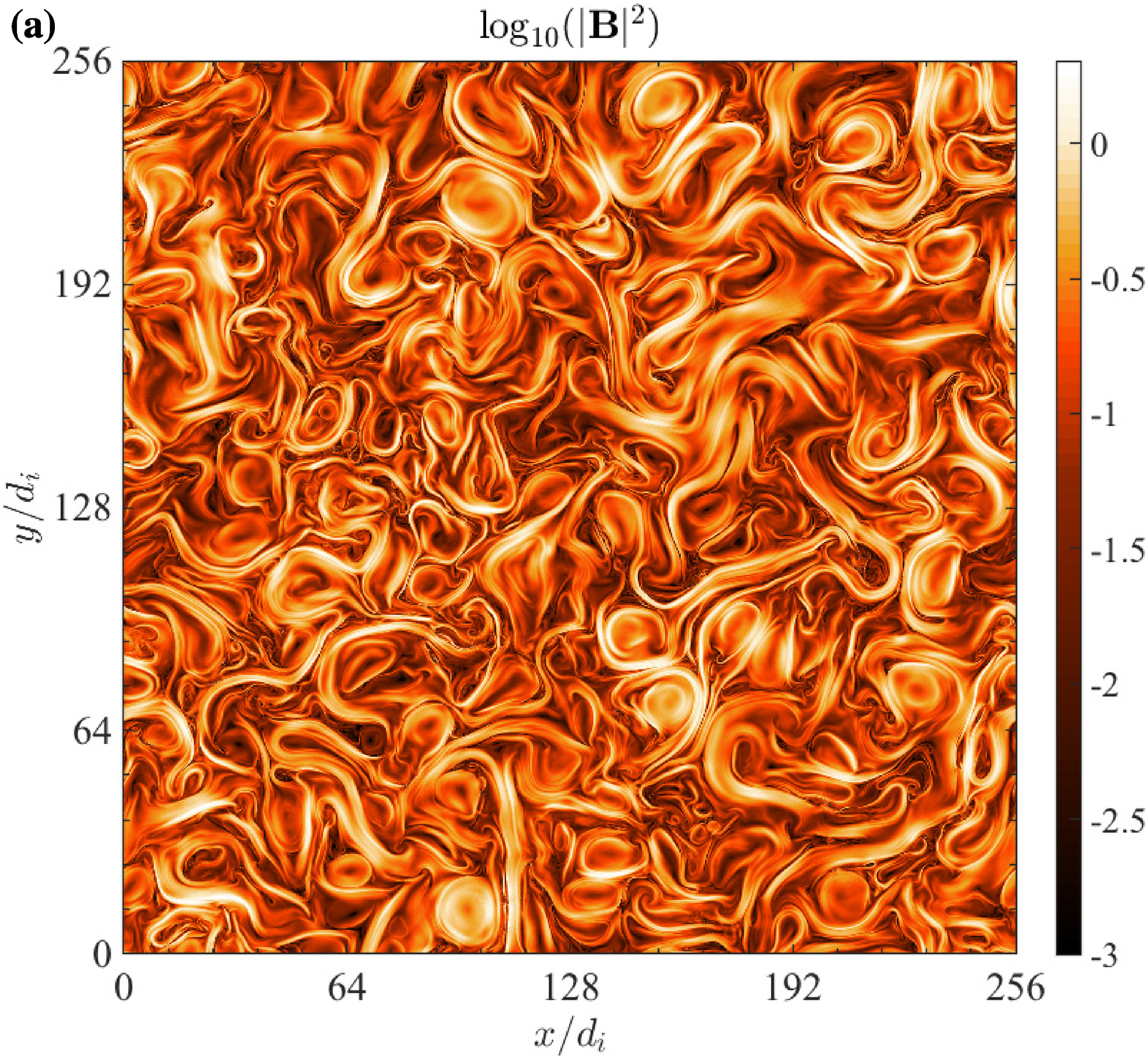}
\includegraphics[width=0.46\textwidth]{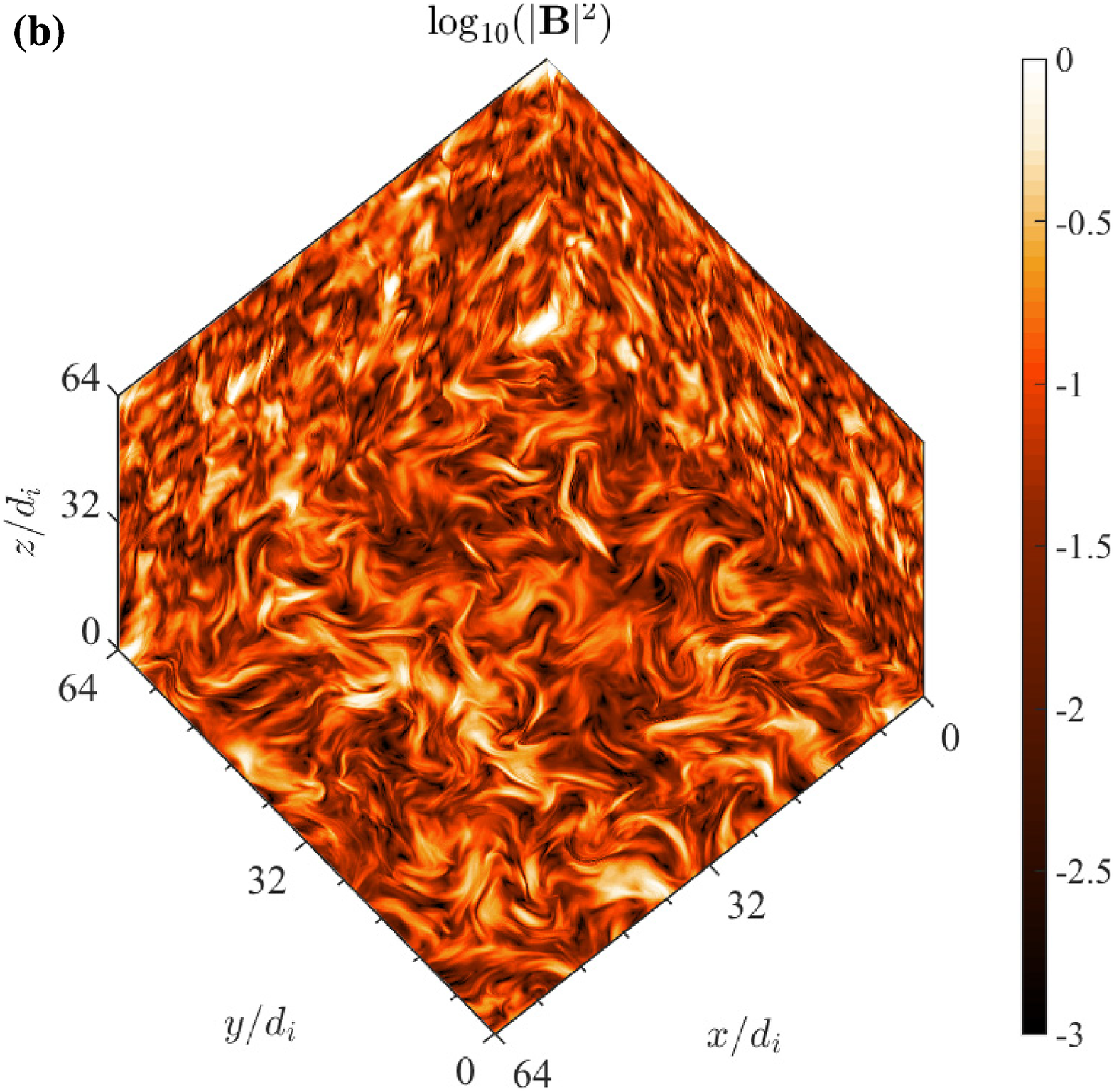}\\
\includegraphics[width=0.48\textwidth]{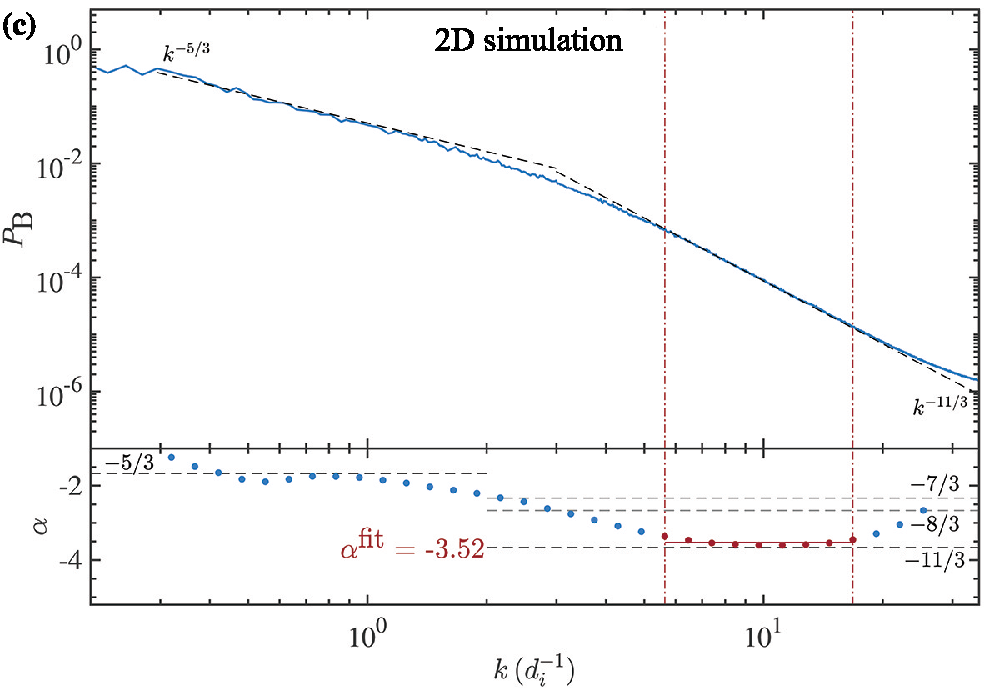}
\includegraphics[width=0.48\textwidth]{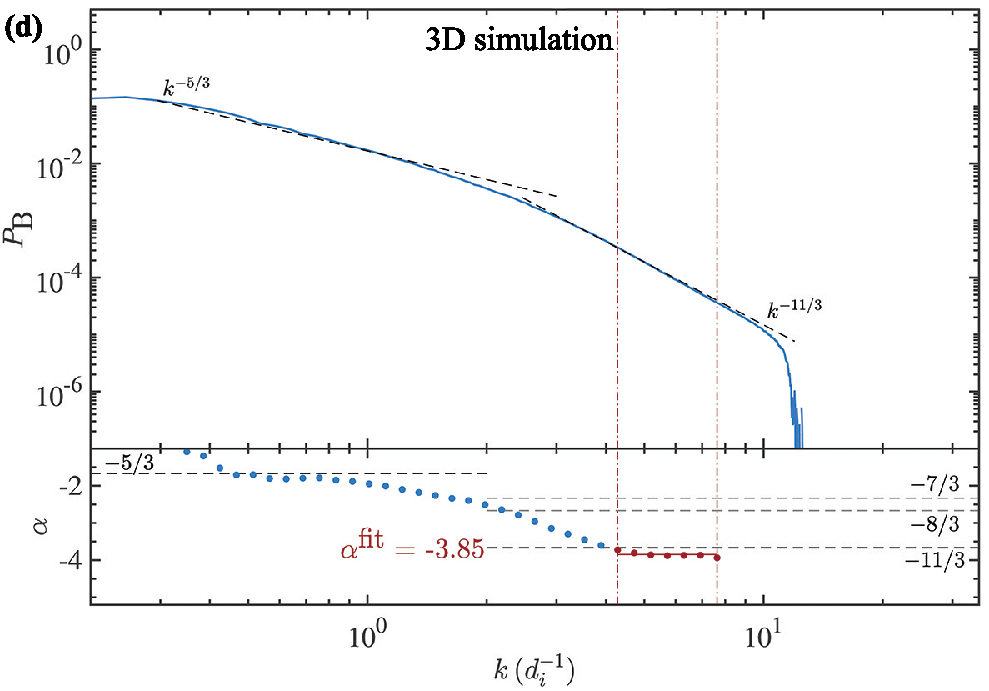}
\caption{Numerical dataset: 2D and 3D hybrid-kinetic simulations of plasma turbulence initialized with values of ion and electron plasma beta similar to those encountered on average at PSP first perihelion, here shown at the time of the maximum turbulent activity. (\textbf{a}) Contour plot of the squared magnitude of the magnetic fluctuations, $|\bm{B}|^2$, for the 2D simulation. The plasma is evolved in a $256\, d_\mathrm{i} \times 256\, d_\mathrm{i}$ domain and the ambient magnetic field $\bm{B}_0 = B_0 \, \hat{z}$ is out of the plane. The system looks qualitatively similar to what have been previously obtained with simulations reproducing plasma conditions typical of the solar wind~\citep{Franci_al_2018b} and of the Earth's magnetopause~\citep{Franci_al_2020}. The main difference is the larger amplitude of the magnetic field gradients, i.e., stronger current sheets, and their larger filling factor. (\textbf{b}) Contour plot of $|\bm{B}|^2$ for the 3D simulation on a $64\, d_\mathrm{i} \times 64\, d_\mathrm{i} \times 64\, d_\mathrm{i}$ domain, with the same physical parameters as the 2D one. Despite the initial fluctuations being isotropic in wavevetor domain, small-scale gradients mainly develop in the $x,y$ plane perpendicular to $\bm{B}_0$ and they look elongated along it, as in previous similar 3D simulations~\citep{Franci_al_2018b}. (\textbf{c}) Isotropized 1D power spectrum of the total magnetic field fluctuations for the 2D simulation. (\textbf{d}) The same as in Fig.~\ref{fig:PSP}c, but for the 3D simulation.}
.\label{fig:DNS}
\end{figure*}

The numerical dataset consists of a 2D and a 3D high-resolution hybrid-kinetic simulations of decaying plasma turbulence performed with the code CAMELIA~\citep{Franci_al_2018a}. Both simulations are initialized with a uniform plasma, embedded in an ambient magnetic field and perturbed with large isotropic Alfv\'enic-like fluctuations, i.e., ion bulk velocity and magnetic fluctuations with the same amplitude in Alfv\'enic units (cf. Sec.~\ref{subsec:methDNS} for all normalization units). The ion and electron plasma betas, i.e., the ratio of particle thermal pressure to the magnetic pressure, were set to be comparable to their average observed values on the day of PSP first perihelion ($\beta_{\mathrm{i}} = 0.2, \, \beta_{\mathrm{e}} = 0.5$), in order to mimic similar plasma conditions.

The same fitting procedure for the observational spectra has been applied to the numerical spectra obtained from the 2D (Fig.~~\ref{fig:DNS}c) and the 3D case (Fig.~\ref{fig:DNS}c). Both numerical spectra show a power-law range at sub-ion scales with a spectral index compatible with $-11/3$, although this is less extended in the latter case due to the larger computational cost which limits the spatial resolution.

\begin{figure}
\includegraphics[width=0.48\textwidth]{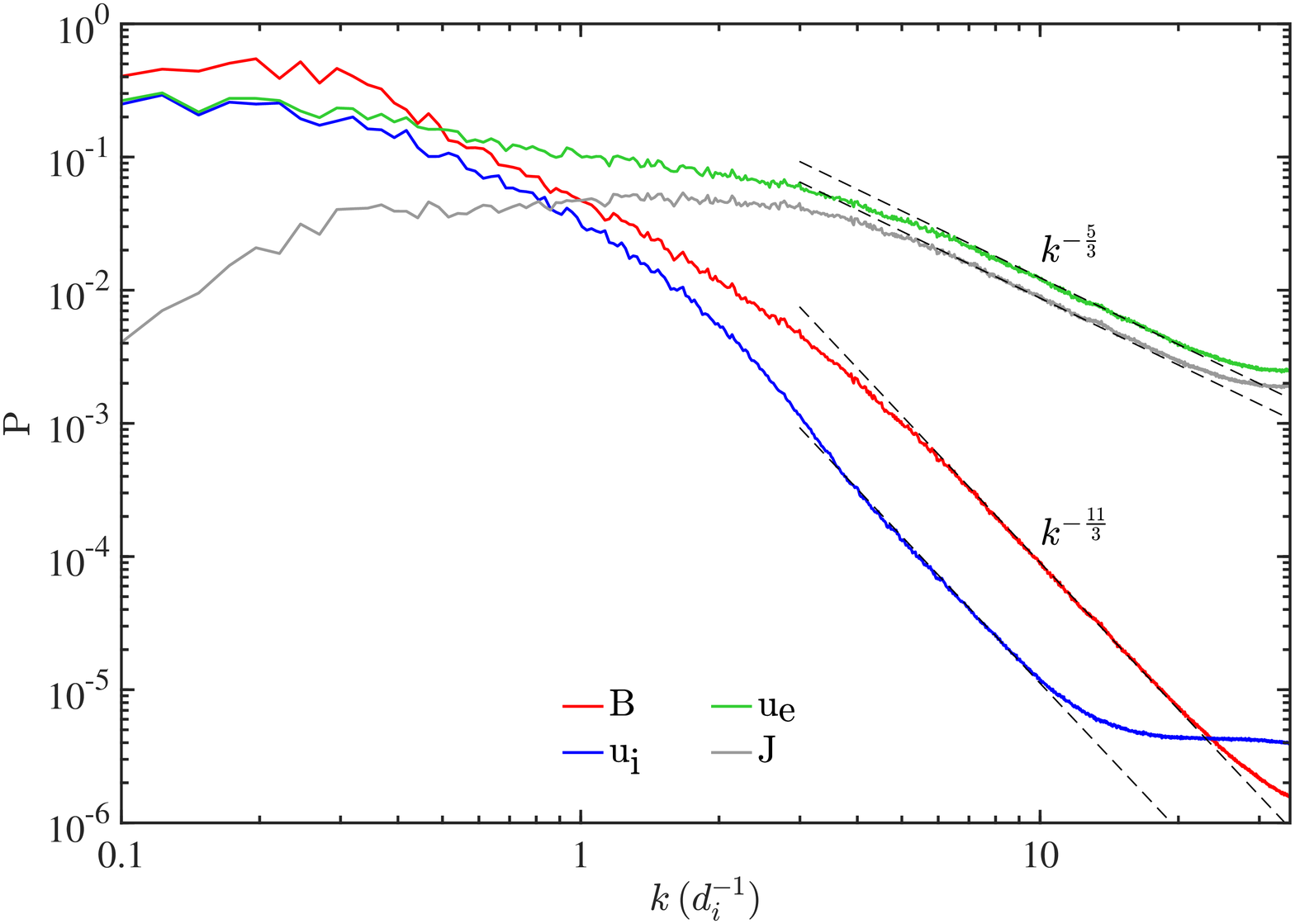}\\
\caption{Power spectra of the magnetic field (red), the ion (blue) and electron (green) bulk velocities, and the current density (grey) for the 2D hybrid-kinetic simulation. Power laws with slope $-5/3$ and $-11/3$ are drawn as a reference at sub-ion scales.}
\label{fig:allspectra}
\end{figure}

Fig.~\ref{fig:allspectra} also shows the spectra of the ion and electron bulk velocities and of the current density for the 2D simulation. The corresponding spectra from the 3D simulation, not shown here, are very similar, especially at $k d_i \gtrsim 1$. 
The power spectrum of the ion bulk velocity fluctuations, $P_{\mathrm{u}_{\mathrm{i}}}$, starts decreasing around $k d_i \gtrsim 1$, consistent with solar wind observations and previous simulations~\citep[e.g.,][]{Franci_al_2018b,Safrankova_al_2016}. It does not, however, drop off steeper than the magnetic fluctuations, as previously observed~\citep{Stawarz_al_2016,Chen_Boldyrev_2017,Franci_al_2020}, but reduces steadily until the noise level is reached. On the contrary, corresponding to the scale of the transition in the magnetic field spectra, it starts behaving as a power law, with a spectral index comparable to that of $P_\mathrm{B}$. An almost constant ratio $P_\mathrm{B}/P_{\mathrm{u}_{\mathrm{i}}} \sim 8$ is observed, i.e., the ion velocity fluctuations are of the same order as the magnetic fluctuations in Alfv\'en units. The ion motion is here still contributing to the plasma dynamics at these scales, providing a non-negligible contribution to the total current density, $\bm{J}$. This is also evident from the fact that the spectra of the current density and of the electron bulk velocity, $\bm{u}_{\mathrm{e}} = \bm{u}_{\mathrm{i}} - \bm{J}/n$ exhibit a slightly different level, although with a similar power-law behavior. The slope is compatible with $-11/3+2 = -5/3$, consistent with $\bm{J} = \nabla \times \bm{B}$, which can be assumed here at all observed scales due to the non-relativistic regime of fluctuations.

\subsubsection{Comparison between observations and simulations}
\begin{figure}
\includegraphics[width=0.47\textwidth]{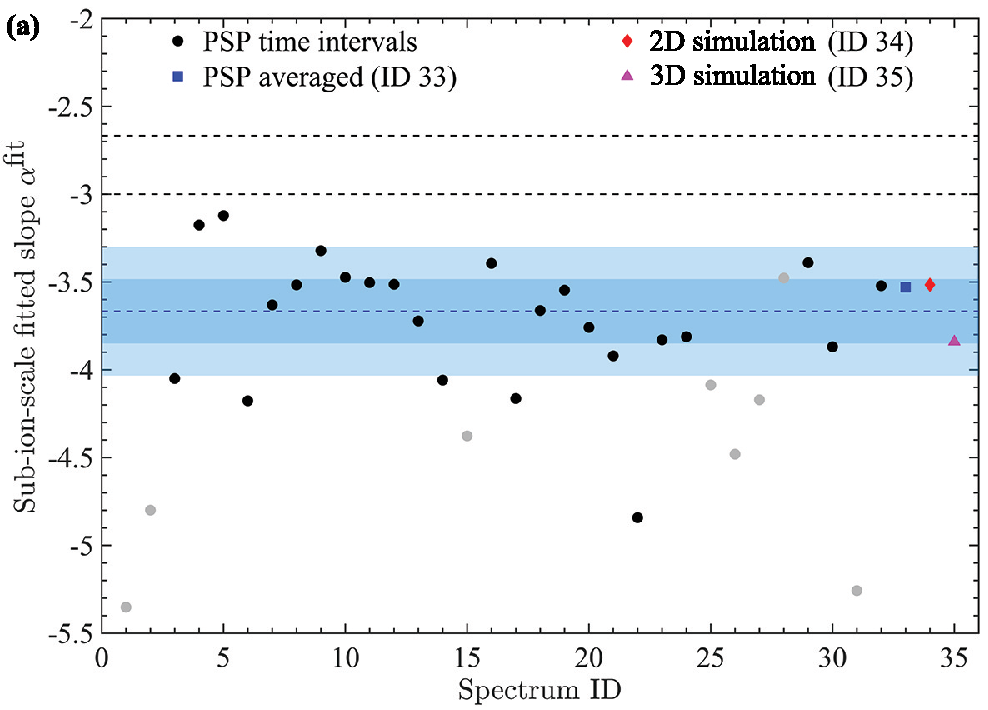}\\
\includegraphics[width=0.48\textwidth]{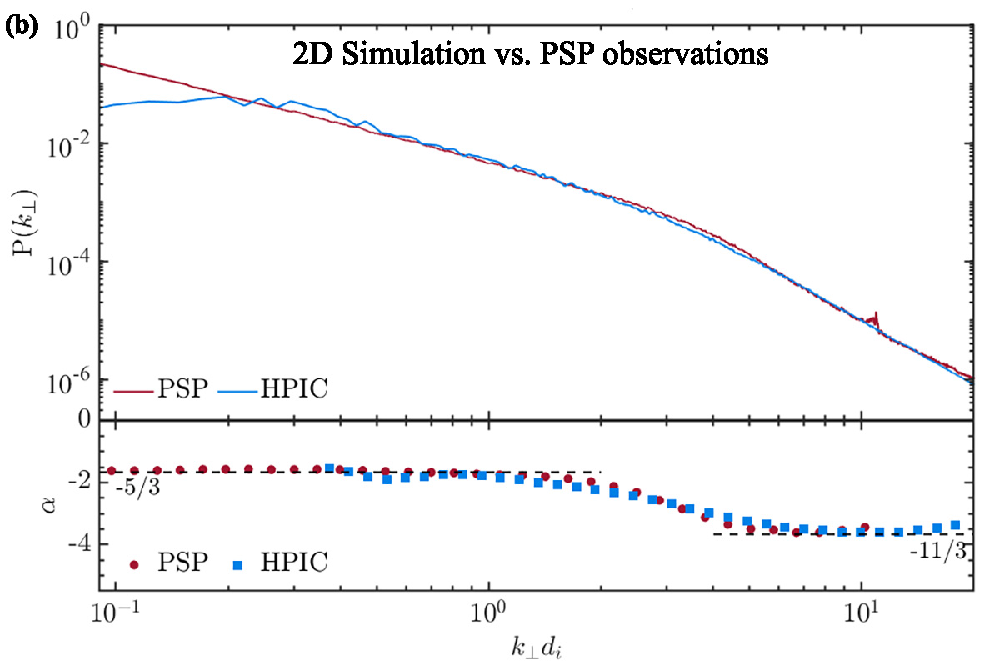}
\caption{Comparison of the spectral properties between observations and simulations. (\textbf{a}) Local spectral index of the magnetic field spectrum at sub-ion scales for each of the 32 45-minuted intervals (IDs from 1 to 32, black points, see Fig.~\ref{fig:PSP}a and Fig.~\ref{fig:PSP}c), for the averaged PSP spectrum (ID 33, blu square, see Fig.~\ref{fig:PSP}d), for the 2D simulation (ID 34, red diamond, see Fig.~\ref{fig:DNS}c), and for the 3D simulation (ID 35, pink triangle, see Fig.~\ref{fig:DNS}d). The grey points mark the intervals where the automatic fit interval has less than a factor of 2 between its lower and upper frequencies. In this case, a power-law behavior is either not present or it's extension is very small and so not significant enough. The horizontal dashed black lines mark the theoretically predicted spectral indices $-7/3$ and $-8/3$, while the blue dashed line marks the prediction of our model, $-11/3$. The two light blue shaded areas mark $5\%$ and $10\%$ around this value. (\textbf{b}) Direct comparison between the observational and the numerical magnetic field spectra as the result of overlapping Fig.~\ref{fig:PSP}d and Fig.~\ref{fig:DNS}c. The former has been arbitrarily shifted along the $x$-axis for the purpose of comparing the slope, with no pretence to use the Taylor hypothesis to accurately convert the frequency in wavenumber.}
\label{fig:slopes}
\end{figure}

The power spectra of the magnetic fluctuations in both observational and numerical data exhibit a power-law behavior at scales just below the ion-scale transition, with slopes compatible with $-11/3$.
Fig.~\ref{fig:slopes}a summarizes and compares the local sub-ion-scale spectral index of all 32 short-interval spectra, of the 24h-averaged PSP spectrum, and of the spectra of the 2D and the 3D hybrid-kinetic simulations. From the distribution of their values, we see that:
i) there are no spectral indices around the theoretically predicted $-7/3$ or $-8/3$ and moreover no interval exhibits a power law flatter than $\sim -3$; ii) most of the intervals have a slope which is very close and compatible with $-11/3$. We can conclude that the $-11/3$ spectral index of the PSP averaged spectrum is not just the result of a few individual spectra dominating over the others. On the contrary, it seems to be the manifestation of a very frequent behavior over the whole day of PSP first perihelion. Fig.~\ref{fig:slopes}b directly compares the observational and the numerical magnetic field spectra, overlapping Fig.~\ref{fig:PSP}d and Fig.~\ref{fig:DNS}c. The former has been arbitrarily shifted along the $x$-axis so that the scale of the transition would coincide, with no pretence to use the Taylor hypothesis to accurately convert the frequency in wavenumber.

\subsubsection{Theoretical interpretation}
\label{sec:model}

Here we provide a theoretical interpretation for the $-11/3$ spectral index in terms of a simple non-relativistic two-fluid model which does not require any ion and electron kinetic processes (e.g., wave-particle resonances) or any electron-inertia effects. 
In the following equations, we will normalize the magnetic field to
the magnitude of the ambient magnetic field, $B_0$,  
the ion and electron bulk velocities to the Alfv\'en velocity, $V_A$, the ion and electron densities to the ambient density, $n_0$, and lengths to the ion inertial length, $d_{\rm i}$. 
We will also use this constraint between the magnetic field and the current density
\begin{equation}
\label{eq:constraint}
\bm{J} = \nabla \times \bm{B} \Longrightarrow  J_k \sim k B_k, 
\end{equation}
which follows from neglecting the displacement current in Amp\`ere's law in the non-relativistic regime. Here and in the following equations, the subscript $k$ denotes the Fourier transform of a quantity. In the limit of small density fluctuations, $n_{\mathrm{i,e}} \sim n_0$, the current density can be written as 
\begin{equation}
\label{eq:current}
\bm{J}_k = \bm{u}_{\mathrm{i},k} - \bm{u}_{\mathrm{e},k}.
\end{equation}
In the same limit, we can express the total energy density at sub-ion scales in normalized units as
\begin{equation}
\label{eq:en_density_approx}
\mathcal{E}_k \approx \frac{1}{2} B_k^{2} + \frac{1}{2}u_{\mathrm{i},k}^{2}.
\end{equation}
The scale-invariant transfer rate is then $\mathcal{E}_k/\tau_{\rm nl}\sim \varepsilon$, where $\tau_{\rm nl}$ is a characteristic nonlinear time expressing the scale-to-scale energy coupling.

\begin{figure}
\includegraphics[width=0.48\textwidth]{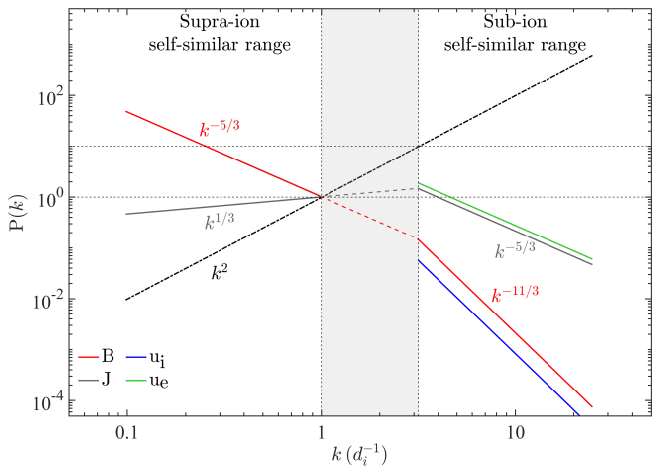}
\caption{Schematic summary of the theoretical model for the plasma regime characterized by power-law magnetic field spectrum with a spectral index of $-11/3$. The main requirement for the regime is that ion bulk velocity fluctuations are not negligible with respect to the magnetic fluctuations, so their spectra are relatively close to each other. As a consequence, the ion motion provides a non-negligible contribution to the current density, which is however still dominated by the electron motion. At sub-ion scales, the spectrum of $\bm{J}$ exhibits a spectral index of $-5/3$ and the one of $\bm{u}_{\mathrm{e}}$ is close to it although not coincident. As a consequence of the constraint expressed by Eq.~\ref{eq:constraint} (black dash-dotted line), this results in a $-11/3$ spectral index for the magnetic field spectrum.}
\label{fig:cartoon}
\end{figure}


At supra-ion scales, 
both observations and simulations indicate that the fluctuations are Alfv\'enic-like, i.e., $u_{\mathrm{i},k}^2 \sim B_k^2$. The nonlinear time associated with the scale-to-scale coupling is therefore $\tau_\mathrm{sup} \sim 1/(k u_{\mathrm{i},k}) \sim 1/(k B_k)$. Demanding a scale-independent energy transfer rate yields the well-known $-5/3$ power-law magnetic field spectrum. Indeed, we observe this power law in both the PSP and the simulation spectra at the largest scales here investigated.

At sub-ion scales, 
Fig.~\ref{fig:allspectra} shows that the ion bulk velocity spectrum follows a power-law behavior which is similar to the one of the magnetic field, although with a lower level. Using Eq.~\ref{eq:constraint}, in the sub-ion range we have
\begin{equation}
\label{eq:requirement}
u_{\mathrm{i},k} \lesssim B_k < k B_k \sim J_k, 
\end{equation}
As a consequence, the current density is mainly due to the electrons although the ions still provide a non-negligible contribution to the plasma dynamics.
Based on this and on the similar spectral behavior of $P_{\mathrm{u}_{\mathrm{e}}}$ and $P_\mathrm{J}$ observed at sub-ion scales (cf. Fig.~\ref{fig:allspectra}), we define the ratio
\begin{equation}
\label{eq:ue_propto_J}
w_k = u_{\mathrm{e},k}/ J_k
\end{equation}
and assume it to be weakly dependent on $k$ in that range, such that it tends to a negative constant of the order of $1$ for sufficiently large $k$, i.e., 
\begin{equation}
\label{eq:wk}
w_k \xrightarrow[k \gg 1]{} - (1 + \epsilon).
\end{equation}
Note that it has been implicitly assumed that $w_k$ is the same for all vector components. This has been verified in the 2D simulation, where we find that $\sqrt{P_{u_{i,\perp}}/P_{J_\perp}} = \sqrt{P_{u_{i,\parallel}}/P_{J_\parallel}} = \sqrt{P_{u_{i}}/P_{J}}$ at scales below the break, which is the range we are interested in.
Using Eq.~\ref{eq:current}, at sub-ion scales we can write the ion bulk velocity fluctuations in terms of the current density as
\begin{equation}
\label{eq:ui_propto_J}
\bm{u}_{\mathrm{i},k} = \bm{J}_k + \bm{u}_{\mathrm{e},k} = (1 + w_k) \bm{J}_k \xrightarrow[k \gg 1]{} - \epsilon \bm{J}_k.
\end{equation}
By substituting Eq.~\ref{eq:ui_propto_J} and Eq.~\ref{eq:constraint} into Eq.~\ref{eq:en_density_approx}, one can express the total energy density in the sub-ion range as a function only of the scale $k$ and of the magnetic fluctuations at such scale through
\begin{equation}
\label{eq:en_density_B}
\mathcal{E}_k \approx \frac{1}{2} ( B_k^2 + \epsilon^2 k^2 B_k^2 ).
\end{equation}
The two terms on the RHS of Eq.~\ref{eq:en_density_B} can now be compared to each other to determine the scaling of the magnetic field spectrum. 

In the limit $\epsilon^2 k^2 \ll 1$, we can approximate the energy at sub-ion scales as $\mathcal{E}_k \approx B_k^2$. We obtain the nonlinear time associated to the evolution of the magnetic field from Faraday's law as  $\tau_\mathrm{nl} \sim (k u_{\mathrm{e},k})^{-1} \sim (k^2 B_k)^{-1}$, where we have used that at sub-ion scales $u_{\mathrm{e},k} \sim J_k = k B_k$. By requesting the energy flux   $\mathcal{E}_k/\tau_\mathrm{nl} \sim k^2 B_k^3$ to be scale-independent, we obtain $P(B) = B_k^2/k \sim k^{-7/3}$. This includes the particular case $\epsilon \rightarrow 0$, when $u_{\mathrm{i},k} \rightarrow 0$ and the current density is all due to the electrons, i.e., the electron-MHD limit~\citep[e.g.,][]{Kingsep_al_1990,Biskamp_al_1996}.

If, on the contrary, 
\begin{equation}
\label{eq:condition11over3}
\epsilon^2 k^2 \gg 1, 
\end{equation} 
we can approximate the energy at sub-ion scales as $\mathcal{E}_k \approx \epsilon^2 k^2 B_k^2$. This term, although coming from the ion kinetic energy, is expressed in terms of the magnetic field via the current density. Therefore, the relevant nonlinear time is again the one associated to the evolution of the magnetic field, i.e., $\tau_\mathrm{nl} \sim (k^2 B_k)^{-1}$. Repeating the same argument as above, we now obtain $P(B) \sim k^{-11/3}$.
The condition expressed by Eq.~\ref{eq:condition11over3} includes the case $\epsilon \equiv -1$, when $u_{\mathrm{e},k} = 0$ and the current is all due to the ions, i.e., ion-MHD~\citep{Meyrand_Galtier_2012}. This, however, represents an extreme case which does not correspond to what we observe in the numerical spectra (see Fig.~\ref{fig:allspectra}).

More in general, we can expect the $-11/3$ power-law regime to occur below a threshold $k_\mathrm{thr} \sim 1/\epsilon$. The threshold will move towards smaller scales the closer $\epsilon$ gets to $0$. In other words, the smaller the ion contribution to the current, the smaller the chances that such regime will occur at scales close to the ion-scale transition.
As a consequence, we should expect the current-mediated regime to be observed at ``large sub-ion scales'' only when the current density is much larger than the ion bulk velocity which, in turn, is not negligible with respect to the magnetic field. 

It is important to note that, although Eq.~\ref{eq:wk} and the last term of Eq.~\ref{eq:ui_propto_J} strictly hold only for $k \gg 1$, we use them in the range $ 3 \lesssim k \lesssim 20$, where the dependence of $w_k$ on $k$ might actually not be negligible. If we assume that such dependence is the form $w_k = - (1 + \epsilon) + a/k + \mathcal{O}(1/k^2)$ (where $a$ is a constant), we obtain $u_{\mathrm{e},k} \sim  (- 1 + a/k  - \epsilon) J_k$ and $u_{\mathrm{i},k} \sim (a/k  - \epsilon) J_k$. At the leading order, this yields $u_{\mathrm{e},k} \sim -J_k$ and $u_{\mathrm{i},k} \sim a J_k/k \sim a B_k$. This would explain why the spectrum of $u_{\mathrm{i},k}$ has a similar slope to that of $B_k$ rather than $J_k$ (cf.  Fig.~\ref{fig:allspectra}). Although we do not investigate here the physical meaning of $w_k$ and of its dependence on $k$, we speculate that this might be somehow related to the alignment between the ion and the electron bulk velocity fluctuations, which could be scale-dependent. 

Fig.~\ref{fig:cartoon} summarizes the main features of the theoretical model. Further details about it can be found in~\citep{Franci_DelSarto_2020}.

All the above arguments has been made in $k$-space. Whenever the Taylor hypothesis holds, however, they can be also applied to the frequency power spectra measured by spacecraft, by using $f=ck$ to link the frequency, $f$, to the wavenumber, $k$, through a constant that depends on the wind speed.

\section{Discussion}
\label{sec:discussion}

This work provides the first combined observational and numerical evidence of a new plasma turbulent regime at scales just below the ion inertial length. This is characterized by: (i) a power-law spectrum of the magnetic field with a spectral index of $-11/3$, and (ii) non-negligible ion bulk velocity fluctuations with respect to the magnetic fluctuations.

We are able to predict and explain such steep spectra with a simple two-fluid model for the turbulent cascade in low-beta plasma that does not include corrections due to intermittency of structures, instabilities, and/or wave- particle resonance, and which is in agreement with both in-situ spacecraft data and hybrid-kinetic simulations.

The dynamics in this regime is driven by the current and therefore by the magnetic field through the constraint expressed by Eq.~\ref{eq:constraint}. This is consistent with the fact that in almost all the time intervals exhibiting the $-11/3$ power-law strong gradients in the magnetic field components are observed, including reversals of the radial component (also called jets or switchbacks). The only two exceptions are intervals \#29 and \#30, which are characterized by an almost constant radial component.
The main requirement, and the key difference with respect to alternative models, is that the ion kinetic energy cannot be neglected in the total energy density, although the ion motion only provides a very small contribution to the total current. 
This regime is therefore very different from previous theoretical models predicting a spectral index of $-11/3$, which typically require one of the following conditions: (i) the kinetic energy overtakes the magnetic energy~\citep{Galtier_Buchlin_2007}, (ii) the current is completely carried by the ions~\citep{Meyrand_Galtier_2012}, or (iii) the ion temperature is much larger that the electron temperature (``inertial kinetic Alfv\`en waves'')~\citep{Chen_Boldyrev_2017}. In our theoretical model none of these represents a necessary condition nor is observed to be satisfied in the numerical simulations presented here. 

Additional simulations with varying physical parameters, currently under investigation, suggest that the combination of the ion plasma beta and the level of magnetic fluctuations with respect to the ambient field play a fundamental role in determining favorable conditions for the current-mediated regime to occur. Both a small beta and a large turbulence amplitude can determine an extended contribution of the ``MHD term'' $\bm{u}_{\mathrm{i}} \times \bm{B}$ to the electric field towards smaller scales below $d_{\mathrm{i}}$ and, as a consequence, to the magnetic field evolution through Faraday's law. Firstly, in a low-beta plasma the ion bulk velocity fluctuations are observed to remain coupled to the magnetic field up to larger values of $k$ (e.g., see Fig. 1 of~\citet{Hellinger_al_2018}). Secondly, if $|\delta\boldsymbol{B}| \sim |\boldsymbol{B}_0|$, then the term $ \delta \bm{u}_{\mathrm{i}} \times \delta \bm{B}$, which is often negligible with respect to $\delta \bm{u}_{\mathrm{i}} \times \bm{B}_0$, will also contribute in Faraday's law. Moreover, it is interesting to note that the contribution from the MHD turbulent cascade flux to the total cascade rate, computed via the third-order structure function, is observed to extend to smaller scales for smaller values of the ion plasma beta (see Fig. 3 of~\citet{Hellinger_al_2018}).

The importance of the small plasma beta in determining a steeper magnetic field spectrum at sub-ion scales has already been hinted in~\citep{Franci_al_2016b}. There, simulations with very small values ($\beta_{\rm{i}}=1/32$ and $1/100$), exhibit a spectral index compatible with $-11/3$ (see the top panel of Fig. 2), despite the much lower level of turbulent fluctuations with respect to the cases presented here ($B^\mathrm{rms}/B_0 \sim 1/8$ and $1/16$, respectively). This suggests that the large-amplitude condition may be necessary only when the plasma beta approaches unity and not for low-beta plasmas.

Our speculation on the role of the plasma beta and the level of fluctuations is also compatible with previous spacecraft observations by \citet[e.g.,][]{Bruno_al_2014}. There, a steepening of the magnetic field spectrum just below $d_{\mathrm{i}}$, with a spectral index compatible with $-11/3$, has been observed in the solar wind, in intervals with lower values of the plasma beta and with a higher power of the fluctuations. This might suggest that the current-mediated regime can be important in other environments rather than just close to the Sun, as long as such conditions are met. The fact that these are more typical of the near-Sun environment, however, might explain why the current-mediated regime can be observed more clearly by PSP and it has not been specifically investigated before in observations.
Both in PSP observations at its first perihelion and in \citep{Bruno_al_2014}, the steep part of the magnetic field spectrum seems to be limited to the range of scales close to $d_{\mathrm{i}}$, i.e., slightly less than a decade in frequency. This is then followed by a flatter part, whose spectral index is more compatible with standard models which predict $-7/3$~\citep{Biskamp_al_1996,Schekochihin_al_2009} or $-8/3$~\citep{Boldyrev_Perez_2012}. Our model provides a simple explanation for this. If the ion velocity fluctuations are not negligible with respect to the magnetic fluctuations at the ion-scale transition, the $-11/3$ regime occurs. If, however, at a smaller scale their contribution to the plasma dynamics become negligible, we recover $\mathcal{E}_k \sim \cancel{u^2_{\mathrm{i}}} + B_k^2$ and therefore the $-7/3$ spectral index. 

A last but not least important aspect to stress is that the current-mediated regime we propose is not a ``kinetic'' regime, in the sense that wave-particle resonance effects, such as ion cyclotron or ion Landau damping, are not necessarily required in order to explain the steepening of the spectrum. Since the dynamic is driven by the current, dissipation is likely rather associated with the disruption of intense current sheets through magnetic reconnection. It is worth mentioning that other features of the simulations presented here (not shown) include enhanced reconnection, enhanced proton heating, and enhanced intermittency. At present, we have not investigated the presence of ion cyclotron waves, which have been detected in PSP observations at its first perihelion~\citep{Bowen_al_2020b,Huang_al_2020}. These are preferentially observed during time intervals characterized by a radial mean magnetic field and cause a bump in the magnetic field spectrum at frequencies corresponding to the ion scales. This is consistent with the results of our analysis: in correspondence with the intervals with a quasi-radial field (grey shaded areas in Fig.~\ref{fig:PSP}b) we do not detect clear power laws in the magnetic field spectrum and this is typically very steep, (see grey dots in Fig.~\ref{fig:slopes}a) compatible with the drop that follows a bump around the ion scales. This further supports the idea that ion cyclotron resonances are unlikely to be responsible for the spectrum with a spectral index of $-11/3$.

Future investigation of the above mentioned aspects will provide a fundamental insight for understanding the role of the energy cascade and current structures on energy dissipation and particle heating in turbulent plasmas. In this regard, it is important to note that we can expect the current-mediated regime to be relevant not only for space plasmas, such as the solar wind and the solar corona, but also for a broad class of other low-beta plasma environments. These can include astrophysical plasmas, e.g., in non-relativistic magnetized jets~\citep{Beskin_Nokhrina_2009} and disks~\citep{Machida_al_2000}, as well as laboratory plasmas, e.g. in magnetic confinement devices if the Hall term in Ohm's law is not negligible~\citep{Zhang_al_2017}, and possibly in turbulent dynamo experiments induced by laser-plasma interactions~\citep{Meinecke_al_2015,Tzeferacos_al_2018}.


\section{Methods}
\subsection{Numerical simulations}
\label{subsec:methDNS}
The numerical dataset was produced using the high-performance 3D hybrid particle-in-cell code CAMELIA \citep{Franci_al_2018b} on the Peta4 machine at the Cambridge Service for Data Driven Discovery (CSD3), within the UK DiRAC infrastructure. Additional preliminary and complementary simulations (not shown here but important as a further confirmation of the results)
were also performed on the Theta HPC system at the Argonne Leadership Computing Facility and on the Marconi HPC system at the Italian SuperComputing centre CINECA. 

CAMELIA is based on the Current Advance Method and Cyclic Leapfrog (CAM-CL) code of Matthews~\citep{Matthews_1994} and solves the Vlasov-Maxwell equations, comprising the equations of motions for individual ions, and the electron fluid equations. The ions are modelled as macroparticles, i.e., fluid elements representing portions of the particles velocity distribution functions in phase space, and are tracked in a
Lagrangian frame, whereas moments of the distribution such as densities and currents are computed simultaneously on Eulerian (stationary) mesh points. Electrons act as a massless, charge-neutralizing fluid. 
The magnetic field is expressed in units of
the magnitude of the ambient magnetic field, $B_0$,  
the ion and electron bulk velocities are in units of the Alfv\'en velocity, $V_A$, time is in units of inverse ion gyrofrequency, $\Omega_{\rm i}^{-1}$, and lengths are in units of the ion inertial length, $d_{\rm i}=v_A/\Omega_{\rm i}$.

The initial condition consists of a homogeneous plasma embedded in a uniform ambient magnetic field, $\bm{B}_0 = B_0 \bm{\hat{z}}$, which in the 2D case is out of the $(x,y)$ simulation plane. We perturb the plasma with Alfv\'enic-like magnetic and ion bulk velocity fluctuations which are perpendicular to $\bm{B}_0$. These are the
 superposition of modes with equal amplitude, random phases, energy equipartition between magnetic and kinetic energy (i.e., with negligible initial residual energy) and they are balanced (i.e., with zero initial cross-helicity).
 Such initialization does not provide a realistic and comprehensive representation of the environment met by PSP at its first perihelion. We try, however, to reproduce similar plasma conditions by setting the ion and electron plasma betas to values compatible with the their average observed value, i.e., $\beta_{\rm{i}}=0.2$ and $\beta_{\rm{e}}=0.5$. The reader can find the detailed time variation of the electron core beta close to the first PSP perihelion in~\citep{Halekas_al_2020}.
The other parameters characterizing the physical initial condition for the 2D and 3D simulations are, respectively: spectrum of initial magnetic fluctuations proportional to $k$ and $k^2$ at $k_\perp^\mathrm{inj} d_i \lesssim 0.4$ and $k^\mathrm{inj} d_i \lesssim 0.5$, 
amplitude of the initial magnetic fluctuations $B^{\mathrm{rms}}/B_0 \sim 0.44$ and $B^{\mathrm{rms}}/B_0 \sim 0.33$.
The numerical setting for the 2D and 3D simulations consists of, respectively: $4096^2$ and $512^3$ grid points, box size $L_{x,y} = 256\,d_i$ and $L_{x,y,z} = 64\,d_i$, spatial resolution
$\delta_{x,y} = d_i/16$ and $\delta_{x,y,z} = d_i/8$, $1024$ and $512$ particle-per-cell, and resistivity $\eta = 1.5 \times 10^{-3}$ and $\eta = 5 \times 10^{-3}$ in units of $4\pi V_A c^{-1}\Omega_i^{-1}$.
All parameters were chosen on the basis of our previous studies, which recovered a qualitative agreement with observations in the solar wind ~\citep[e.g.,][]{Franci_al_2015a,Franci_al_2016b} and a quantitative agreement with observation at the Earth's magnetopause~\citep{Franci_al_2020}. Futher details about our numerical setup and its implementation can be found in~\citet{Franci_al_2015b} and~\citep{Franci_al_2018a} for the 2D and the 3D simulations, respectively.
Simulation data are here analysed and shown at $t=50 \; \Omega_{\rm i}^{-1}$, i.e., the time when the RMS value of the current density, $\bm{J}$, reached its maximum. This is universally regarded as a proxy for the development of a fully-developed quasi-stationary turbulent state. We have verified that the spectral properties do not change significantly around that time. The time evolution of global
quantities (e.g., the rms values of $\bm{J}$ and of the magnetic
fluctuations, $\bm{B}$, not shown here) is qualitatively the same as
in previous 2D~\citep{Franci_al_2015b} and 3D~\citep{Franci_al_2018b} hybrid-kinetic simulations, the quantitative difference being the time needed for the maximum turbulent activity to develop. This is inversely proportional to the initial rms of fluctuations, $B^{\mathrm{rms}}$, and to the injection-scale wavenumber, $k^{\mathrm{inj}}$, through the non-linear time $\tau_{\mathrm{nl}} \sim (k^{\mathrm{inj}} B^{\mathrm{rms}})^{-1}$. 
This work represents a further evidence that a 2D simulation may be representative enough of the spectral behavior in a more realistic 3D geometry, due to the quasi-2D nature of the turbulent cascade~\citep[e.g.,][]{Franci_al_2018b,Franci_al_2020}.
Based on this, we only show and analyse here the spectra from the 2D simulation, which employs a much higher spatial resolutions and therefore exhibits more extended power laws. 

\subsection{Spacecraft observations}
\label{subsec:methPSP}
The in-situ spacecraft data shown in Fig.~\ref{fig:PSP} have been obtained from the FIELDS Fluxgate Magnetometer (MAG) onboard Parker-Solar-Probe at its first perihelion (from 00:00 UT to 23:59 UT on 6 November 2018). The data are in the RTN (radial-tangential-normal) reference frame, which is centered at the spacecraft and oriented with respect to the line connecting it to the Sun: the R axis is directed radially away from the Sun, the T axis is the cross product of the Sun's spin vector and the R axis, and the N axis completes the right handed set.
The time series for the three components of the magnetic field have been divided in 32 subintervals of 45mins.
For each time interval, the time series have been normalized to the amplitude of their respective mean vector magnetic field. The corresponding multitaper power spectral densities have been computed using 16 tapers.

\subsection{Power spectra and local the spectral index}
\label{subsec:methFIT}
For the simulations, we define the 1D omnidirectional power spectrum of a quantity $\Psi(\bm{x})$ as
\begin{equation}
P_\Psi(k)\equiv \delta\hat{\Psi}^2(k)/k,
\end{equation}
where $\delta\hat{\Psi}(k)$ is the amplitude of the fluctuations of $\Psi$ at the
scale corresponding to $k$. In the main text, we will omit both the $\delta$ and the hat symbol and express the fluctuations $\delta\hat{\Psi}^2(k)$ just as $\Psi_k$.\\
In the 2D case, this is obtained by integrating over circles in the 2D plane $(k_x,k_y)$ delimited by $k = |\bm{k}_\bot| = \sqrt{k_x^2+k_y^2}$ and $k+dk$. Due to the numerical discretization of the grid, the integral is replaced with the sum 
\begin{equation}
P_\Psi(k)=\sum_{\sqrt{k_x^2+k_y^2} =k}|\hat{\Psi}|^2(k_x,k_y),
\end{equation}
where $\hat{\Psi}$ are the (complex) Fourier coefficients. \\
In the 3D case, this is obtained by integrating over spherical shells in the 3D space $(k_x,k_y,k_z)$ delimited by $k = |\bm{k}| = \sqrt{k_x^2+k_y^2+k_z^2}$ and $k+dk$, which corresponds to \begin{equation}
P_\Psi(k)=\sum_{\sqrt{k_x^2+k_y^2+k_z^2} = k}|\hat{\Psi}|^2(k_x,k_y,k_z).
\end{equation}
For observations, the power spectrum of a quantity $\Psi(t)$ is 
\begin{equation}
P_\Psi(f)\equiv\delta\hat{\Psi}^2(f),
\end{equation}
where $\delta \hat{\Psi}(f)$ is the amplitude of the fluctuations of $\Psi$ at the frequency $f$.\\
The local spectral index, $\alpha$, is computing by fitting  magnetic field spectrum over a sliding window of half a decade in wavenumber for the simulations and in frequency for observations. Once this is computed in the whole range of scale, its value at sub-ion scales is obtained through an automatic fitting procedure, where the fit interval is chosen automatically including the minimum value and all the surrounding points within $10\%$ from that value. This automatic fitting procedure is applied to the PSP spectra of all 32 time intervals, to the 24h-averaged PSP spectrum, and to the 2D and the 3D numerical spectra.

\section{Data availability}
The observational data used in this study are available at the NASA Space Physics Data Facility (SPDF), https://spdf.gsfc.nasa.gov/index.html. \emph{The numerical data will be uploaded to EUDAT (https://b2share.eudat.eu/) and made publicly available before the manuscript is published, as soon as the title and the journal are definitive. This will allow us to include those details in the description of the dataset. This section will be then updated accordingly, including the corresponding doi of the dataset.}
\\

\section{References}
\bibliography{references}

\begin{thebibliography}{46}%
\makeatletter
\providecommand \@ifxundefined [1]{%
 \@ifx{#1\undefined}
}%
\providecommand \@ifnum [1]{%
 \ifnum #1\expandafter \@firstoftwo
 \else \expandafter \@secondoftwo
 \fi
}%
\providecommand \@ifx [1]{%
 \ifx #1\expandafter \@firstoftwo
 \else \expandafter \@secondoftwo
 \fi
}%
\providecommand \natexlab [1]{#1}%
\providecommand \enquote  [1]{``#1''}%
\providecommand \bibnamefont  [1]{#1}%
\providecommand \bibfnamefont [1]{#1}%
\providecommand \citenamefont [1]{#1}%
\providecommand \href@noop [0]{\@secondoftwo}%
\providecommand \href [0]{\begingroup \@sanitize@url \@href}%
\providecommand \@href[1]{\@@startlink{#1}\@@href}%
\providecommand \@@href[1]{\endgroup#1\@@endlink}%
\providecommand \@sanitize@url [0]{\catcode `\\12\catcode `\$12\catcode
  `\&12\catcode `\#12\catcode `\^12\catcode `\_12\catcode `\%12\relax}%
\providecommand \@@startlink[1]{}%
\providecommand \@@endlink[0]{}%
\providecommand \url  [0]{\begingroup\@sanitize@url \@url }%
\providecommand \@url [1]{\endgroup\@href {#1}{\urlprefix }}%
\providecommand \urlprefix  [0]{URL }%
\providecommand \Eprint [0]{\href }%
\providecommand \doibase [0]{https://doi.org/}%
\providecommand \selectlanguage [0]{\@gobble}%
\providecommand \bibinfo  [0]{\@secondoftwo}%
\providecommand \bibfield  [0]{\@secondoftwo}%
\providecommand \translation [1]{[#1]}%
\providecommand \BibitemOpen [0]{}%
\providecommand \bibitemStop [0]{}%
\providecommand \bibitemNoStop [0]{.\EOS\space}%
\providecommand \EOS [0]{\spacefactor3000\relax}%
\providecommand \BibitemShut  [1]{\csname bibitem#1\endcsname}%
\let\auto@bib@innerbib\@empty
\bibitem [{\citenamefont {{Tu}}\ and\ \citenamefont
  {{Marsch}}(1995)}]{Tu_Marsch_1995}%
  \BibitemOpen
  \bibfield  {author} {\bibinfo {author} {\bibfnamefont {C.-Y.}\ \bibnamefont
  {{Tu}}}\ and\ \bibinfo {author} {\bibfnamefont {E.}~\bibnamefont
  {{Marsch}}},\ }\bibfield  {title} {\bibinfo {title} {{MHD structures, waves
  and turbulence in the solar wind: Observations and theories}},\ }\href
  {https://doi.org/10.1007/BF00748891} {\bibfield  {journal} {\bibinfo
  {journal} {Space Sci.~Rev.}\ }\textbf {\bibinfo {volume} {73}},\ \bibinfo
  {pages} {1} (\bibinfo {year} {1995})}\BibitemShut {NoStop}%
\bibitem [{\citenamefont {{Matthaeus}}\ and\ \citenamefont
  {{Velli}}(2011)}]{Matthaeus_Velli_2011}%
  \BibitemOpen
  \bibfield  {author} {\bibinfo {author} {\bibfnamefont {W.~H.}\ \bibnamefont
  {{Matthaeus}}}\ and\ \bibinfo {author} {\bibfnamefont {M.}~\bibnamefont
  {{Velli}}},\ }\bibfield  {title} {\bibinfo {title} {{Who Needs Turbulence?. A
  Review of Turbulence Effects in the Heliosphere and on the Fundamental
  Process of Reconnection}},\ }\href
  {https://doi.org/10.1007/s11214-011-9793-9} {\bibfield  {journal} {\bibinfo
  {journal} {Space Sci.~Rev.}\ }\textbf {\bibinfo {volume} {160}},\ \bibinfo
  {pages} {145} (\bibinfo {year} {2011})}\BibitemShut {NoStop}%
\bibitem [{\citenamefont {{Alexandrova}}\ \emph {et~al.}(2013)\citenamefont
  {{Alexandrova}}, \citenamefont {{Chen}}, \citenamefont {{Sorriso-Valvo}},
  \citenamefont {{Horbury}},\ and\ \citenamefont
  {{Bale}}}]{Alexandrova_al_2013}%
  \BibitemOpen
  \bibfield  {author} {\bibinfo {author} {\bibfnamefont {O.}~\bibnamefont
  {{Alexandrova}}}, \bibinfo {author} {\bibfnamefont {C.~H.~K.}\ \bibnamefont
  {{Chen}}}, \bibinfo {author} {\bibfnamefont {L.}~\bibnamefont
  {{Sorriso-Valvo}}}, \bibinfo {author} {\bibfnamefont {T.~S.}\ \bibnamefont
  {{Horbury}}},\ and\ \bibinfo {author} {\bibfnamefont {S.~D.}\ \bibnamefont
  {{Bale}}},\ }\bibfield  {title} {\bibinfo {title} {{Solar Wind Turbulence and
  the Role of Ion Instabilities}},\ }\href
  {https://doi.org/10.1007/s11214-013-0004-8} {\bibfield  {journal} {\bibinfo
  {journal} {Space Sci.~Rev.}\ }\textbf {\bibinfo {volume} {178}},\ \bibinfo
  {pages} {101} (\bibinfo {year} {2013})}\BibitemShut {NoStop}%
\bibitem [{\citenamefont {{Bruno}}\ and\ \citenamefont
  {{Carbone}}(2013)}]{Bruno_Carbone_2013}%
  \BibitemOpen
  \bibfield  {author} {\bibinfo {author} {\bibfnamefont {R.}~\bibnamefont
  {{Bruno}}}\ and\ \bibinfo {author} {\bibfnamefont {V.}~\bibnamefont
  {{Carbone}}},\ }\bibfield  {title} {\bibinfo {title} {{The Solar Wind as a
  Turbulence Laboratory}},\ }\href {https://doi.org/10.12942/lrsp-2013-2}
  {\bibfield  {journal} {\bibinfo  {journal} {Living Rev. Solar Phys.}\
  }\textbf {\bibinfo {volume} {10}},\ \bibinfo {pages} {2} (\bibinfo {year}
  {2013})}\BibitemShut {NoStop}%
\bibitem [{\citenamefont {{Sahraoui}}\ \emph {et~al.}(2009)\citenamefont
  {{Sahraoui}}, \citenamefont {{Goldstein}}, \citenamefont {{Robert}},\ and\
  \citenamefont {{Khotyaintsev}}}]{Sahraoui_al_2009}%
  \BibitemOpen
  \bibfield  {author} {\bibinfo {author} {\bibfnamefont {F.}~\bibnamefont
  {{Sahraoui}}}, \bibinfo {author} {\bibfnamefont {M.~L.}\ \bibnamefont
  {{Goldstein}}}, \bibinfo {author} {\bibfnamefont {P.}~\bibnamefont
  {{Robert}}},\ and\ \bibinfo {author} {\bibfnamefont {Y.~V.}\ \bibnamefont
  {{Khotyaintsev}}},\ }\bibfield  {title} {\bibinfo {title} {{Evidence of a
  Cascade and Dissipation of Solar-Wind Turbulence at the Electron
  Gyroscale}},\ }\href {https://doi.org/10.1103/PhysRevLett.102.231102}
  {\bibfield  {journal} {\bibinfo  {journal} {Phys.~Rev.~Lett.}\ }\textbf
  {\bibinfo {volume} {102}},\ \bibinfo {eid} {231102} (\bibinfo {year}
  {2009})}\BibitemShut {NoStop}%
\bibitem [{\citenamefont {{Kiyani}}\ \emph {et~al.}(2015)\citenamefont
  {{Kiyani}}, \citenamefont {{Osman}},\ and\ \citenamefont
  {{Chapman}}}]{Kiyani_al_2015}%
  \BibitemOpen
  \bibfield  {author} {\bibinfo {author} {\bibfnamefont {K.~H.}\ \bibnamefont
  {{Kiyani}}}, \bibinfo {author} {\bibfnamefont {K.~T.}\ \bibnamefont
  {{Osman}}},\ and\ \bibinfo {author} {\bibfnamefont {S.~C.}\ \bibnamefont
  {{Chapman}}},\ }\bibfield  {title} {\bibinfo {title} {{Dissipation and
  heating in solar wind turbulence: from the macro to the micro and back
  again}},\ }\href {https://doi.org/10.1098/rsta.2014.0155} {\bibfield
  {journal} {\bibinfo  {journal} {Philos. Trans. Royal Soc. A}\ }\textbf
  {\bibinfo {volume} {373}},\ \bibinfo {pages} {20140155} (\bibinfo {year}
  {2015})}\BibitemShut {NoStop}%
\bibitem [{\citenamefont {{Podesta}}\ \emph {et~al.}(2007)\citenamefont
  {{Podesta}}, \citenamefont {{Roberts}},\ and\ \citenamefont
  {{Goldstein}}}]{Podesta_al_2007}%
  \BibitemOpen
  \bibfield  {author} {\bibinfo {author} {\bibfnamefont {J.~J.}\ \bibnamefont
  {{Podesta}}}, \bibinfo {author} {\bibfnamefont {D.~A.}\ \bibnamefont
  {{Roberts}}},\ and\ \bibinfo {author} {\bibfnamefont {M.~L.}\ \bibnamefont
  {{Goldstein}}},\ }\bibfield  {title} {\bibinfo {title} {{Spectral Exponents
  of Kinetic and Magnetic Energy Spectra in Solar Wind Turbulence}},\ }\href
  {https://doi.org/10.1086/519211} {\bibfield  {journal} {\bibinfo  {journal}
  {Astrophys.~J.}\ }\textbf {\bibinfo {volume} {664}},\ \bibinfo {pages} {543}
  (\bibinfo {year} {2007})}\BibitemShut {NoStop}%
\bibitem [{\citenamefont {Marsch}(2004)}]{Marsch2004}%
  \BibitemOpen
  \bibfield  {author} {\bibinfo {author} {\bibfnamefont {E.}~\bibnamefont
  {Marsch}},\ }\bibinfo {title} {Waves and turbulence in the solar corona},\
  in\ \href {https://doi.org/10.1007/978-1-4020-2831-1_10} {\emph {\bibinfo
  {booktitle} {The Sun and the Heliosphere as an Integrated System}}},\
  \bibinfo {editor} {edited by\ \bibinfo {editor} {\bibfnamefont
  {G.}~\bibnamefont {Poletto}}\ and\ \bibinfo {editor} {\bibfnamefont {S.~T.}\
  \bibnamefont {Suess}}}\ (\bibinfo  {publisher} {Springer Netherlands},\
  \bibinfo {address} {Dordrecht},\ \bibinfo {year} {2004})\ pp.\ \bibinfo
  {pages} {283--317}\BibitemShut {NoStop}%
\bibitem [{\citenamefont {Balbus}\ and\ \citenamefont
  {Hawley}(1998)}]{Balbus_Hawley_1998}%
  \BibitemOpen
  \bibfield  {author} {\bibinfo {author} {\bibfnamefont {S.~A.}\ \bibnamefont
  {Balbus}}\ and\ \bibinfo {author} {\bibfnamefont {J.~F.}\ \bibnamefont
  {Hawley}},\ }\bibfield  {title} {\bibinfo {title} {Instability, turbulence,
  and enhanced transport in accretion disks},\ }\href
  {https://doi.org/10.1103/RevModPhys.70.1} {\bibfield  {journal} {\bibinfo
  {journal} {Rev. Mod. Phys.}\ }\textbf {\bibinfo {volume} {70}},\ \bibinfo
  {pages} {1} (\bibinfo {year} {1998})}\BibitemShut {NoStop}%
\bibitem [{\citenamefont {Ferri{\`{e}}re}(2019)}]{Ferriere_2019}%
  \BibitemOpen
  \bibfield  {author} {\bibinfo {author} {\bibfnamefont {K.}~\bibnamefont
  {Ferri{\`{e}}re}},\ }\bibfield  {title} {\bibinfo {title} {Plasma turbulence
  in the interstellar medium},\ }\href
  {https://doi.org/10.1088/1361-6587/ab49eb} {\bibfield  {journal} {\bibinfo
  {journal} {Plasma Phys. Control. Fusion}\ }\textbf {\bibinfo {volume} {62}},\
  \bibinfo {pages} {014014} (\bibinfo {year} {2019})}\BibitemShut {NoStop}%
\bibitem [{\citenamefont {Schekochihin}\ and\ \citenamefont
  {Cowley}(2006)}]{Schekochihin_Cowley_2006}%
  \BibitemOpen
  \bibfield  {author} {\bibinfo {author} {\bibfnamefont {A.~A.}\ \bibnamefont
  {Schekochihin}}\ and\ \bibinfo {author} {\bibfnamefont {S.~C.}\ \bibnamefont
  {Cowley}},\ }\bibfield  {title} {\bibinfo {title} {Turbulence, magnetic
  fields, and plasma physics in clusters of galaxies},\ }\href
  {https://doi.org/10.1063/1.2179053} {\bibfield  {journal} {\bibinfo
  {journal} {Phys. Plasmas}\ }\textbf {\bibinfo {volume} {13}},\ \bibinfo
  {pages} {056501} (\bibinfo {year} {2006})},\ \Eprint
  {https://arxiv.org/abs/https://doi.org/10.1063/1.2179053}
  {https://doi.org/10.1063/1.2179053} \BibitemShut {NoStop}%
\bibitem [{\citenamefont {Fox}\ \emph {et~al.}(2016)\citenamefont {Fox},
  \citenamefont {Velli}, \citenamefont {Bale}, \citenamefont {Decker},
  \citenamefont {Driesman}, \citenamefont {Howard}, \citenamefont {Kasper},
  \citenamefont {Kinnison}, \citenamefont {Kusterer}, \citenamefont {Lario},
  \citenamefont {Lockwood}, \citenamefont {McComas}, \citenamefont {Raouafi},\
  and\ \citenamefont {Szabo}}]{Fox_al_2016}%
  \BibitemOpen
  \bibfield  {author} {\bibinfo {author} {\bibfnamefont {N.~J.}\ \bibnamefont
  {Fox}}, \bibinfo {author} {\bibfnamefont {M.~C.}\ \bibnamefont {Velli}},
  \bibinfo {author} {\bibfnamefont {S.~D.}\ \bibnamefont {Bale}}, \bibinfo
  {author} {\bibfnamefont {R.}~\bibnamefont {Decker}}, \bibinfo {author}
  {\bibfnamefont {A.}~\bibnamefont {Driesman}}, \bibinfo {author}
  {\bibfnamefont {R.~A.}\ \bibnamefont {Howard}}, \bibinfo {author}
  {\bibfnamefont {J.~C.}\ \bibnamefont {Kasper}}, \bibinfo {author}
  {\bibfnamefont {J.}~\bibnamefont {Kinnison}}, \bibinfo {author}
  {\bibfnamefont {M.}~\bibnamefont {Kusterer}}, \bibinfo {author}
  {\bibfnamefont {D.}~\bibnamefont {Lario}}, \bibinfo {author} {\bibfnamefont
  {M.~K.}\ \bibnamefont {Lockwood}}, \bibinfo {author} {\bibfnamefont {D.~J.}\
  \bibnamefont {McComas}}, \bibinfo {author} {\bibfnamefont {N.~E.}\
  \bibnamefont {Raouafi}},\ and\ \bibinfo {author} {\bibfnamefont
  {A.}~\bibnamefont {Szabo}},\ }\bibfield  {title} {\bibinfo {title} {The solar
  probe plus mission: Humanity's first visit to our star},\ }\href
  {https://doi.org/10.1007/s11214-015-0211-6} {\bibfield  {journal} {\bibinfo
  {journal} {Space Sci. Rev.}\ }\textbf {\bibinfo {volume} {204}},\ \bibinfo
  {pages} {7} (\bibinfo {year} {2016})}\BibitemShut {NoStop}%
\bibitem [{\citenamefont {Bale}\ \emph {et~al.}(2019)\citenamefont {Bale},
  \citenamefont {Badman}, \citenamefont {Bonnell}, \citenamefont {Bowen},
  \citenamefont {Burgess}, \citenamefont {Case}, \citenamefont {Cattell},
  \citenamefont {Chandran}, \citenamefont {Chaston}, \citenamefont {Chen},
  \citenamefont {Drake}, \citenamefont {{de Wit}}, \citenamefont {Eastwood},
  \citenamefont {Ergun}, \citenamefont {Farrell}, \citenamefont {Fong},
  \citenamefont {Goetz}, \citenamefont {Goldstein}, \citenamefont {Goodrich},
  \citenamefont {Harvey}, \citenamefont {Horbury}, \citenamefont {Howes},
  \citenamefont {Kasper}, \citenamefont {Kellogg}, \citenamefont {Klimchuk},
  \citenamefont {Korreck}, \citenamefont {Krasnoselskikh}, \citenamefont
  {Krucker}, \citenamefont {Laker}, \citenamefont {Larson}, \citenamefont
  {MacDowall}, \citenamefont {Maksimovic}, \citenamefont {Malaspina},
  \citenamefont {Martinez-Oliveros}, \citenamefont {McComas}, \citenamefont
  {Meyer-Vernet}, \citenamefont {Moncuquet}, \citenamefont {Mozer},
  \citenamefont {Phan}, \citenamefont {Pulupa}, \citenamefont {Raouafi},
  \citenamefont {Salem}, \citenamefont {Stansby}, \citenamefont {Stevens},
  \citenamefont {Szabo}, \citenamefont {Velli}, \citenamefont {Woolley},\ and\
  \citenamefont {Wygant}}]{Bale_al_2019}%
  \BibitemOpen
  \bibfield  {author} {\bibinfo {author} {\bibfnamefont {S.}~\bibnamefont
  {Bale}}, \bibinfo {author} {\bibfnamefont {S.}~\bibnamefont {Badman}},
  \bibinfo {author} {\bibfnamefont {J.}~\bibnamefont {Bonnell}}, \bibinfo
  {author} {\bibfnamefont {T.}~\bibnamefont {Bowen}}, \bibinfo {author}
  {\bibfnamefont {D.}~\bibnamefont {Burgess}}, \bibinfo {author} {\bibfnamefont
  {A.}~\bibnamefont {Case}}, \bibinfo {author} {\bibfnamefont {C.}~\bibnamefont
  {Cattell}}, \bibinfo {author} {\bibfnamefont {B.}~\bibnamefont {Chandran}},
  \bibinfo {author} {\bibfnamefont {C.}~\bibnamefont {Chaston}}, \bibinfo
  {author} {\bibfnamefont {C.}~\bibnamefont {Chen}}, \bibinfo {author}
  {\bibfnamefont {J.}~\bibnamefont {Drake}}, \bibinfo {author} {\bibfnamefont
  {T.}~\bibnamefont {{de Wit}}}, \bibinfo {author} {\bibfnamefont
  {J.}~\bibnamefont {Eastwood}}, \bibinfo {author} {\bibfnamefont
  {R.}~\bibnamefont {Ergun}}, \bibinfo {author} {\bibfnamefont
  {W.}~\bibnamefont {Farrell}}, \bibinfo {author} {\bibfnamefont
  {C.}~\bibnamefont {Fong}}, \bibinfo {author} {\bibfnamefont {K.}~\bibnamefont
  {Goetz}}, \bibinfo {author} {\bibfnamefont {M.}~\bibnamefont {Goldstein}},
  \bibinfo {author} {\bibfnamefont {K.}~\bibnamefont {Goodrich}}, \bibinfo
  {author} {\bibfnamefont {P.}~\bibnamefont {Harvey}}, \bibinfo {author}
  {\bibfnamefont {T.}~\bibnamefont {Horbury}}, \bibinfo {author} {\bibfnamefont
  {G.}~\bibnamefont {Howes}}, \bibinfo {author} {\bibfnamefont
  {J.}~\bibnamefont {Kasper}}, \bibinfo {author} {\bibfnamefont
  {P.}~\bibnamefont {Kellogg}}, \bibinfo {author} {\bibfnamefont
  {J.}~\bibnamefont {Klimchuk}}, \bibinfo {author} {\bibfnamefont
  {K.}~\bibnamefont {Korreck}}, \bibinfo {author} {\bibfnamefont
  {V.}~\bibnamefont {Krasnoselskikh}}, \bibinfo {author} {\bibfnamefont
  {S.}~\bibnamefont {Krucker}}, \bibinfo {author} {\bibfnamefont
  {R.}~\bibnamefont {Laker}}, \bibinfo {author} {\bibfnamefont
  {D.}~\bibnamefont {Larson}}, \bibinfo {author} {\bibfnamefont
  {R.}~\bibnamefont {MacDowall}}, \bibinfo {author} {\bibfnamefont
  {M.}~\bibnamefont {Maksimovic}}, \bibinfo {author} {\bibfnamefont
  {D.}~\bibnamefont {Malaspina}}, \bibinfo {author} {\bibfnamefont
  {J.}~\bibnamefont {Martinez-Oliveros}}, \bibinfo {author} {\bibfnamefont
  {D.}~\bibnamefont {McComas}}, \bibinfo {author} {\bibfnamefont
  {N.}~\bibnamefont {Meyer-Vernet}}, \bibinfo {author} {\bibfnamefont
  {M.}~\bibnamefont {Moncuquet}}, \bibinfo {author} {\bibfnamefont
  {F.}~\bibnamefont {Mozer}}, \bibinfo {author} {\bibfnamefont
  {T.}~\bibnamefont {Phan}}, \bibinfo {author} {\bibfnamefont {M.}~\bibnamefont
  {Pulupa}}, \bibinfo {author} {\bibfnamefont {N.}~\bibnamefont {Raouafi}},
  \bibinfo {author} {\bibfnamefont {C.}~\bibnamefont {Salem}}, \bibinfo
  {author} {\bibfnamefont {D.}~\bibnamefont {Stansby}}, \bibinfo {author}
  {\bibfnamefont {M.}~\bibnamefont {Stevens}}, \bibinfo {author} {\bibfnamefont
  {A.}~\bibnamefont {Szabo}}, \bibinfo {author} {\bibfnamefont
  {M.}~\bibnamefont {Velli}}, \bibinfo {author} {\bibfnamefont
  {T.}~\bibnamefont {Woolley}},\ and\ \bibinfo {author} {\bibfnamefont
  {J.}~\bibnamefont {Wygant}},\ }\bibfield  {title} {\bibinfo {title} {Highly
  structured slow solar wind emerging from an equatorial coronal hole},\ }\href
  {https://doi.org/10.1038/s41586-019-1818-7} {\bibfield  {journal} {\bibinfo
  {journal} {Nature}\ }\textbf {\bibinfo {volume} {576}},\ \bibinfo {pages}
  {237} (\bibinfo {year} {2019})}\BibitemShut {NoStop}%
\bibitem [{\citenamefont {de~Wit}\ \emph {et~al.}(2020)\citenamefont {de~Wit},
  \citenamefont {Krasnoselskikh}, \citenamefont {Bale}, \citenamefont
  {Bonnell}, \citenamefont {Bowen}, \citenamefont {Chen}, \citenamefont
  {Froment}, \citenamefont {Goetz}, \citenamefont {Harvey}, \citenamefont
  {Jagarlamudi}, \citenamefont {Larosa}, \citenamefont {MacDowall},
  \citenamefont {Malaspina}, \citenamefont {Matthaeus}, \citenamefont {Pulupa},
  \citenamefont {Velli},\ and\ \citenamefont {Whittlesey}}]{Dudok_de_Wit_2020}%
  \BibitemOpen
  \bibfield  {author} {\bibinfo {author} {\bibfnamefont {T.~D.}\ \bibnamefont
  {de~Wit}}, \bibinfo {author} {\bibfnamefont {V.~V.}\ \bibnamefont
  {Krasnoselskikh}}, \bibinfo {author} {\bibfnamefont {S.~D.}\ \bibnamefont
  {Bale}}, \bibinfo {author} {\bibfnamefont {J.~W.}\ \bibnamefont {Bonnell}},
  \bibinfo {author} {\bibfnamefont {T.~A.}\ \bibnamefont {Bowen}}, \bibinfo
  {author} {\bibfnamefont {C.~H.~K.}\ \bibnamefont {Chen}}, \bibinfo {author}
  {\bibfnamefont {C.}~\bibnamefont {Froment}}, \bibinfo {author} {\bibfnamefont
  {K.}~\bibnamefont {Goetz}}, \bibinfo {author} {\bibfnamefont {P.~R.}\
  \bibnamefont {Harvey}}, \bibinfo {author} {\bibfnamefont {V.~K.}\
  \bibnamefont {Jagarlamudi}}, \bibinfo {author} {\bibfnamefont
  {A.}~\bibnamefont {Larosa}}, \bibinfo {author} {\bibfnamefont {R.~J.}\
  \bibnamefont {MacDowall}}, \bibinfo {author} {\bibfnamefont {D.~M.}\
  \bibnamefont {Malaspina}}, \bibinfo {author} {\bibfnamefont {W.~H.}\
  \bibnamefont {Matthaeus}}, \bibinfo {author} {\bibfnamefont {M.}~\bibnamefont
  {Pulupa}}, \bibinfo {author} {\bibfnamefont {M.}~\bibnamefont {Velli}},\ and\
  \bibinfo {author} {\bibfnamefont {P.~L.}\ \bibnamefont {Whittlesey}},\
  }\bibfield  {title} {\bibinfo {title} {Switchbacks in the near-sun magnetic
  field: Long memory and impact on the turbulence cascade},\ }\href
  {https://doi.org/10.3847/1538-4365/ab5853} {\bibfield  {journal} {\bibinfo
  {journal} {Astrophys. J., Suppl. Ser.}\ }\textbf {\bibinfo {volume} {246}},\
  \bibinfo {pages} {39} (\bibinfo {year} {2020})}\BibitemShut {NoStop}%
\bibitem [{\citenamefont {Bowen}\ \emph {et~al.}(2020)\citenamefont {Bowen},
  \citenamefont {Mallet}, \citenamefont {Huang}, \citenamefont {Klein},
  \citenamefont {Malaspina}, \citenamefont {Stevens}, \citenamefont {Bale},
  \citenamefont {Bonnell}, \citenamefont {Case}, \citenamefont {Chandran},\
  and\ \citenamefont {et~al.}}]{Bowen_al_2020b}%
  \BibitemOpen
  \bibfield  {author} {\bibinfo {author} {\bibfnamefont {T.~A.}\ \bibnamefont
  {Bowen}}, \bibinfo {author} {\bibfnamefont {A.}~\bibnamefont {Mallet}},
  \bibinfo {author} {\bibfnamefont {J.}~\bibnamefont {Huang}}, \bibinfo
  {author} {\bibfnamefont {K.~G.}\ \bibnamefont {Klein}}, \bibinfo {author}
  {\bibfnamefont {D.~M.}\ \bibnamefont {Malaspina}}, \bibinfo {author}
  {\bibfnamefont {M.}~\bibnamefont {Stevens}}, \bibinfo {author} {\bibfnamefont
  {S.~D.}\ \bibnamefont {Bale}}, \bibinfo {author} {\bibfnamefont {J.~W.}\
  \bibnamefont {Bonnell}}, \bibinfo {author} {\bibfnamefont {A.~W.}\
  \bibnamefont {Case}}, \bibinfo {author} {\bibfnamefont {B.~D.~G.}\
  \bibnamefont {Chandran}},\ and\ \bibinfo {author} {\bibnamefont {et~al.}},\
  }\bibfield  {title} {\bibinfo {title} {Ion-scale electromagnetic waves in the
  inner heliosphere},\ }\href {https://doi.org/10.3847/1538-4365/ab6c65}
  {\bibfield  {journal} {\bibinfo  {journal} {Astrophys. J., Suppl. Ser.}\
  }\textbf {\bibinfo {volume} {246}},\ \bibinfo {pages} {66} (\bibinfo {year}
  {2020})}\BibitemShut {NoStop}%
\bibitem [{\citenamefont {Huang}\ \emph {et~al.}(2020)\citenamefont {Huang},
  \citenamefont {Zhang}, \citenamefont {Sahraoui}, \citenamefont {He},
  \citenamefont {Yuan}, \citenamefont {Andr{\'{e}}s}, \citenamefont {Hadid},
  \citenamefont {Deng}, \citenamefont {Jiang}, \citenamefont {Yu},
  \citenamefont {Xiong}, \citenamefont {Wei}, \citenamefont {Xu}, \citenamefont
  {Bale},\ and\ \citenamefont {Kasper}}]{Huang_al_2020}%
  \BibitemOpen
  \bibfield  {author} {\bibinfo {author} {\bibfnamefont {S.~Y.}\ \bibnamefont
  {Huang}}, \bibinfo {author} {\bibfnamefont {J.}~\bibnamefont {Zhang}},
  \bibinfo {author} {\bibfnamefont {F.}~\bibnamefont {Sahraoui}}, \bibinfo
  {author} {\bibfnamefont {J.~S.}\ \bibnamefont {He}}, \bibinfo {author}
  {\bibfnamefont {Z.~G.}\ \bibnamefont {Yuan}}, \bibinfo {author}
  {\bibfnamefont {N.}~\bibnamefont {Andr{\'{e}}s}}, \bibinfo {author}
  {\bibfnamefont {L.~Z.}\ \bibnamefont {Hadid}}, \bibinfo {author}
  {\bibfnamefont {X.~H.}\ \bibnamefont {Deng}}, \bibinfo {author}
  {\bibfnamefont {K.}~\bibnamefont {Jiang}}, \bibinfo {author} {\bibfnamefont
  {L.}~\bibnamefont {Yu}}, \bibinfo {author} {\bibfnamefont {Q.~Y.}\
  \bibnamefont {Xiong}}, \bibinfo {author} {\bibfnamefont {Y.~Y.}\ \bibnamefont
  {Wei}}, \bibinfo {author} {\bibfnamefont {S.~B.}\ \bibnamefont {Xu}},
  \bibinfo {author} {\bibfnamefont {S.~D.}\ \bibnamefont {Bale}},\ and\
  \bibinfo {author} {\bibfnamefont {J.~C.}\ \bibnamefont {Kasper}},\ }\bibfield
   {title} {\bibinfo {title} {Kinetic scale slow solar wind turbulence in the
  inner heliosphere: Coexistence of kinetic alfv{\'{e}}n waves and alfv{\'{e}}n
  ion cyclotron waves},\ }\href {https://doi.org/10.3847/2041-8213/ab9abb}
  {\bibfield  {journal} {\bibinfo  {journal} {Astrophys. J.}\ }\textbf
  {\bibinfo {volume} {897}},\ \bibinfo {pages} {L3} (\bibinfo {year}
  {2020})}\BibitemShut {NoStop}%
\bibitem [{\citenamefont {Bandyopadhyay}\ \emph {et~al.}(2020)\citenamefont
  {Bandyopadhyay}, \citenamefont {Goldstein}, \citenamefont {Maruca},
  \citenamefont {Matthaeus}, \citenamefont {Parashar}, \citenamefont {Ruffolo},
  \citenamefont {Chhiber}, \citenamefont {Usmanov}, \citenamefont {Chasapis},
  \citenamefont {Qudsi}, \citenamefont {Bale}, \citenamefont {Bonnell},
  \citenamefont {de~Wit}, \citenamefont {Goetz}, \citenamefont {Harvey},
  \citenamefont {MacDowall}, \citenamefont {Malaspina}, \citenamefont {Pulupa},
  \citenamefont {Kasper}, \citenamefont {Korreck}, \citenamefont {Case},
  \citenamefont {Stevens}, \citenamefont {Whittlesey}, \citenamefont {Larson},
  \citenamefont {Livi}, \citenamefont {Klein}, \citenamefont {Velli},\ and\
  \citenamefont {Raouafi}}]{Bandyopadhyay_al_2020b}%
  \BibitemOpen
  \bibfield  {author} {\bibinfo {author} {\bibfnamefont {R.}~\bibnamefont
  {Bandyopadhyay}}, \bibinfo {author} {\bibfnamefont {M.~L.}\ \bibnamefont
  {Goldstein}}, \bibinfo {author} {\bibfnamefont {B.~A.}\ \bibnamefont
  {Maruca}}, \bibinfo {author} {\bibfnamefont {W.~H.}\ \bibnamefont
  {Matthaeus}}, \bibinfo {author} {\bibfnamefont {T.~N.}\ \bibnamefont
  {Parashar}}, \bibinfo {author} {\bibfnamefont {D.}~\bibnamefont {Ruffolo}},
  \bibinfo {author} {\bibfnamefont {R.}~\bibnamefont {Chhiber}}, \bibinfo
  {author} {\bibfnamefont {A.}~\bibnamefont {Usmanov}}, \bibinfo {author}
  {\bibfnamefont {A.}~\bibnamefont {Chasapis}}, \bibinfo {author}
  {\bibfnamefont {R.}~\bibnamefont {Qudsi}}, \bibinfo {author} {\bibfnamefont
  {S.~D.}\ \bibnamefont {Bale}}, \bibinfo {author} {\bibfnamefont {J.~W.}\
  \bibnamefont {Bonnell}}, \bibinfo {author} {\bibfnamefont {T.~D.}\
  \bibnamefont {de~Wit}}, \bibinfo {author} {\bibfnamefont {K.}~\bibnamefont
  {Goetz}}, \bibinfo {author} {\bibfnamefont {P.~R.}\ \bibnamefont {Harvey}},
  \bibinfo {author} {\bibfnamefont {R.~J.}\ \bibnamefont {MacDowall}}, \bibinfo
  {author} {\bibfnamefont {D.~M.}\ \bibnamefont {Malaspina}}, \bibinfo {author}
  {\bibfnamefont {M.}~\bibnamefont {Pulupa}}, \bibinfo {author} {\bibfnamefont
  {J.~C.}\ \bibnamefont {Kasper}}, \bibinfo {author} {\bibfnamefont {K.~E.}\
  \bibnamefont {Korreck}}, \bibinfo {author} {\bibfnamefont {A.~W.}\
  \bibnamefont {Case}}, \bibinfo {author} {\bibfnamefont {M.}~\bibnamefont
  {Stevens}}, \bibinfo {author} {\bibfnamefont {P.}~\bibnamefont {Whittlesey}},
  \bibinfo {author} {\bibfnamefont {D.}~\bibnamefont {Larson}}, \bibinfo
  {author} {\bibfnamefont {R.}~\bibnamefont {Livi}}, \bibinfo {author}
  {\bibfnamefont {K.~G.}\ \bibnamefont {Klein}}, \bibinfo {author}
  {\bibfnamefont {M.}~\bibnamefont {Velli}},\ and\ \bibinfo {author}
  {\bibfnamefont {N.}~\bibnamefont {Raouafi}},\ }\bibfield  {title} {\bibinfo
  {title} {Enhanced energy transfer rate in solar wind turbulence observed near
  the sun from parker solar probe},\ }\href
  {https://doi.org/10.3847/1538-4365/ab5dae} {\bibfield  {journal} {\bibinfo
  {journal} {Astrophys. J., Suppl. Ser.}\ }\textbf {\bibinfo {volume} {246}},\
  \bibinfo {pages} {48} (\bibinfo {year} {2020})}\BibitemShut {NoStop}%
\bibitem [{\citenamefont {Martinovi{\'{c}}}\ \emph {et~al.}(2020)\citenamefont
  {Martinovi{\'{c}}}, \citenamefont {Klein}, \citenamefont {Kasper},
  \citenamefont {Case}, \citenamefont {Korreck}, \citenamefont {Larson},
  \citenamefont {Livi}, \citenamefont {Stevens}, \citenamefont {Whittlesey},
  \citenamefont {Chandran}, \citenamefont {Alterman}, \citenamefont {Huang},
  \citenamefont {Chen}, \citenamefont {Bale}, \citenamefont {Pulupa},
  \citenamefont {Malaspina}, \citenamefont {Bonnell}, \citenamefont {Harvey},
  \citenamefont {Goetz}, \citenamefont {de~Wit},\ and\ \citenamefont
  {MacDowall}}]{Martinovic_al_2020}%
  \BibitemOpen
  \bibfield  {author} {\bibinfo {author} {\bibfnamefont {M.~M.}\ \bibnamefont
  {Martinovi{\'{c}}}}, \bibinfo {author} {\bibfnamefont {K.~G.}\ \bibnamefont
  {Klein}}, \bibinfo {author} {\bibfnamefont {J.~C.}\ \bibnamefont {Kasper}},
  \bibinfo {author} {\bibfnamefont {A.~W.}\ \bibnamefont {Case}}, \bibinfo
  {author} {\bibfnamefont {K.~E.}\ \bibnamefont {Korreck}}, \bibinfo {author}
  {\bibfnamefont {D.}~\bibnamefont {Larson}}, \bibinfo {author} {\bibfnamefont
  {R.}~\bibnamefont {Livi}}, \bibinfo {author} {\bibfnamefont {M.}~\bibnamefont
  {Stevens}}, \bibinfo {author} {\bibfnamefont {P.}~\bibnamefont {Whittlesey}},
  \bibinfo {author} {\bibfnamefont {B.~D.~G.}\ \bibnamefont {Chandran}},
  \bibinfo {author} {\bibfnamefont {B.~L.}\ \bibnamefont {Alterman}}, \bibinfo
  {author} {\bibfnamefont {J.}~\bibnamefont {Huang}}, \bibinfo {author}
  {\bibfnamefont {C.~H.~K.}\ \bibnamefont {Chen}}, \bibinfo {author}
  {\bibfnamefont {S.~D.}\ \bibnamefont {Bale}}, \bibinfo {author}
  {\bibfnamefont {M.}~\bibnamefont {Pulupa}}, \bibinfo {author} {\bibfnamefont
  {D.~M.}\ \bibnamefont {Malaspina}}, \bibinfo {author} {\bibfnamefont {J.~W.}\
  \bibnamefont {Bonnell}}, \bibinfo {author} {\bibfnamefont {P.~R.}\
  \bibnamefont {Harvey}}, \bibinfo {author} {\bibfnamefont {K.}~\bibnamefont
  {Goetz}}, \bibinfo {author} {\bibfnamefont {T.~D.}\ \bibnamefont {de~Wit}},\
  and\ \bibinfo {author} {\bibfnamefont {R.~J.}\ \bibnamefont {MacDowall}},\
  }\bibfield  {title} {\bibinfo {title} {The enhancement of proton stochastic
  heating in the near-sun solar wind},\ }\href
  {https://doi.org/10.3847/1538-4365/ab527f} {\bibfield  {journal} {\bibinfo
  {journal} {Astrophys. J., Suppl. Ser.}\ }\textbf {\bibinfo {volume} {246}},\
  \bibinfo {pages} {30} (\bibinfo {year} {2020})}\BibitemShut {NoStop}%
\bibitem [{\citenamefont {{Bowen}}\ \emph {et~al.}(2020)\citenamefont
  {{Bowen}}, \citenamefont {{Mallet}}, \citenamefont {{Bale}}, \citenamefont
  {{Bonnell}}, \citenamefont {{Case}}, \citenamefont {{Chandran}},
  \citenamefont {{Chasapis}}, \citenamefont {{Chen}}, \citenamefont {{Duan}},
  \citenamefont {{Dudok de Wit}}, \citenamefont {{Goetz}}, \citenamefont
  {{Halekas}}, \citenamefont {{Harvey}}, \citenamefont {{Kasper}},
  \citenamefont {{Korreck}}, \citenamefont {{Larson}}, \citenamefont {{Livi}},
  \citenamefont {{MacDowall}}, \citenamefont {{Malaspina}}, \citenamefont
  {{McManus}}, \citenamefont {{Pulupa}}, \citenamefont {{Stevens}},\ and\
  \citenamefont {{Whittlesey}}}]{Bowen_al_2020c}%
  \BibitemOpen
  \bibfield  {author} {\bibinfo {author} {\bibfnamefont {T.~A.}\ \bibnamefont
  {{Bowen}}}, \bibinfo {author} {\bibfnamefont {A.}~\bibnamefont {{Mallet}}},
  \bibinfo {author} {\bibfnamefont {S.~D.}\ \bibnamefont {{Bale}}}, \bibinfo
  {author} {\bibfnamefont {J.~W.}\ \bibnamefont {{Bonnell}}}, \bibinfo {author}
  {\bibfnamefont {A.~W.}\ \bibnamefont {{Case}}}, \bibinfo {author}
  {\bibfnamefont {B.~D.~G.}\ \bibnamefont {{Chandran}}}, \bibinfo {author}
  {\bibfnamefont {A.}~\bibnamefont {{Chasapis}}}, \bibinfo {author}
  {\bibfnamefont {C.~H.~K.}\ \bibnamefont {{Chen}}}, \bibinfo {author}
  {\bibfnamefont {D.}~\bibnamefont {{Duan}}}, \bibinfo {author} {\bibfnamefont
  {T.}~\bibnamefont {{Dudok de Wit}}}, \bibinfo {author} {\bibfnamefont
  {K.}~\bibnamefont {{Goetz}}}, \bibinfo {author} {\bibfnamefont {J.~S.}\
  \bibnamefont {{Halekas}}}, \bibinfo {author} {\bibfnamefont {P.~R.}\
  \bibnamefont {{Harvey}}}, \bibinfo {author} {\bibfnamefont {J.~C.}\
  \bibnamefont {{Kasper}}}, \bibinfo {author} {\bibfnamefont {K.~E.}\
  \bibnamefont {{Korreck}}}, \bibinfo {author} {\bibfnamefont {D.}~\bibnamefont
  {{Larson}}}, \bibinfo {author} {\bibfnamefont {R.}~\bibnamefont {{Livi}}},
  \bibinfo {author} {\bibfnamefont {R.~J.}\ \bibnamefont {{MacDowall}}},
  \bibinfo {author} {\bibfnamefont {D.~M.}\ \bibnamefont {{Malaspina}}},
  \bibinfo {author} {\bibfnamefont {M.~D.}\ \bibnamefont {{McManus}}}, \bibinfo
  {author} {\bibfnamefont {M.}~\bibnamefont {{Pulupa}}}, \bibinfo {author}
  {\bibfnamefont {M.}~\bibnamefont {{Stevens}}},\ and\ \bibinfo {author}
  {\bibfnamefont {P.}~\bibnamefont {{Whittlesey}}},\ }\bibfield  {title}
  {\bibinfo {title} {{Constraining Ion-Scale Heating and Spectral Energy
  Transfer in Observations of Plasma Turbulence}},\ }\href
  {https://doi.org/10.1103/PhysRevLett.125.025102} {\bibfield  {journal}
  {\bibinfo  {journal} {Phys.~Rev.~Lett.}\ }\textbf {\bibinfo {volume} {125}},\
  \bibinfo {eid} {025102} (\bibinfo {year} {2020})}\BibitemShut {NoStop}%
\bibitem [{\citenamefont {{Schekochihin}}\ \emph {et~al.}(2009)\citenamefont
  {{Schekochihin}}, \citenamefont {{Cowley}}, \citenamefont {{Dorland}},
  \citenamefont {{Hammett}}, \citenamefont {{Howes}}, \citenamefont
  {{Quataert}},\ and\ \citenamefont {{Tatsuno}}}]{Schekochihin_al_2009}%
  \BibitemOpen
  \bibfield  {author} {\bibinfo {author} {\bibfnamefont {A.~A.}\ \bibnamefont
  {{Schekochihin}}}, \bibinfo {author} {\bibfnamefont {S.~C.}\ \bibnamefont
  {{Cowley}}}, \bibinfo {author} {\bibfnamefont {W.}~\bibnamefont {{Dorland}}},
  \bibinfo {author} {\bibfnamefont {G.~W.}\ \bibnamefont {{Hammett}}}, \bibinfo
  {author} {\bibfnamefont {G.~G.}\ \bibnamefont {{Howes}}}, \bibinfo {author}
  {\bibfnamefont {E.}~\bibnamefont {{Quataert}}},\ and\ \bibinfo {author}
  {\bibfnamefont {T.}~\bibnamefont {{Tatsuno}}},\ }\bibfield  {title} {\bibinfo
  {title} {{Astrophysical Gyrokinetics: Kinetic and Fluid Turbulent Cascades in
  Magnetized Weakly Collisional Plasmas}},\ }\href
  {https://doi.org/10.1088/0067-0049/182/1/310} {\bibfield  {journal} {\bibinfo
   {journal} {Astrophys. J. Supl. Series}\ }\textbf {\bibinfo {volume} {182}},\
  \bibinfo {pages} {310} (\bibinfo {year} {2009})}\BibitemShut {NoStop}%
\bibitem [{\citenamefont {{Boldyrev}}\ and\ \citenamefont
  {{Perez}}(2012)}]{Boldyrev_Perez_2012}%
  \BibitemOpen
  \bibfield  {author} {\bibinfo {author} {\bibfnamefont {S.}~\bibnamefont
  {{Boldyrev}}}\ and\ \bibinfo {author} {\bibfnamefont {J.~C.}\ \bibnamefont
  {{Perez}}},\ }\bibfield  {title} {\bibinfo {title} {{Spectrum of
  Kinetic-Alfv{\'e}n Turbulence}},\ }\href
  {https://doi.org/10.1088/2041-8205/758/2/L44} {\bibfield  {journal} {\bibinfo
   {journal} {Astrophys.~J.~Lett.}\ }\textbf {\bibinfo {volume} {758}},\
  \bibinfo {eid} {L44} (\bibinfo {year} {2012})}\BibitemShut {NoStop}%
\bibitem [{\citenamefont {{Bale}}\ \emph {et~al.}(2016)\citenamefont {{Bale}},
  \citenamefont {{Goetz}}, \citenamefont {{Harvey}}, \citenamefont {{Turin}},
  \citenamefont {{Bonnell}}, \citenamefont {{Dudok de Wit}}, \citenamefont
  {{Ergun}}, \citenamefont {{MacDowall}}, \citenamefont {{Pulupa}},
  \citenamefont {{Andre}}, \citenamefont {{Bolton}}, \citenamefont
  {{Bougeret}}, \citenamefont {{Bowen}}, \citenamefont {{Burgess}},
  \citenamefont {{Cattell}}, \citenamefont {{Chandran}}, \citenamefont
  {{Chaston}}, \citenamefont {{Chen}}, \citenamefont {{Choi}}, \citenamefont
  {{Connerney}}, \citenamefont {{Cranmer}}, \citenamefont {{Diaz-Aguado}},
  \citenamefont {{Donakowski}}, \citenamefont {{Drake}}, \citenamefont
  {{Farrell}}, \citenamefont {{Fergeau}}, \citenamefont {{Fermin}},
  \citenamefont {{Fischer}}, \citenamefont {{Fox}}, \citenamefont {{Glaser}},
  \citenamefont {{Goldstein}}, \citenamefont {{Gordon}}, \citenamefont
  {{Hanson}}, \citenamefont {{Harris}}, \citenamefont {{Hayes}}, \citenamefont
  {{Hinze}}, \citenamefont {{Hollweg}}, \citenamefont {{Horbury}},
  \citenamefont {{Howard}}, \citenamefont {{Hoxie}}, \citenamefont {{Jannet}},
  \citenamefont {{Karlsson}}, \citenamefont {{Kasper}}, \citenamefont
  {{Kellogg}}, \citenamefont {{Kien}}, \citenamefont {{Klimchuk}},
  \citenamefont {{Krasnoselskikh}}, \citenamefont {{Krucker}}, \citenamefont
  {{Lynch}}, \citenamefont {{Maksimovic}}, \citenamefont {{Malaspina}},
  \citenamefont {{Marker}}, \citenamefont {{Martin}}, \citenamefont
  {{Martinez-Oliveros}}, \citenamefont {{McCauley}}, \citenamefont {{McComas}},
  \citenamefont {{McDonald}}, \citenamefont {{Meyer-Vernet}}, \citenamefont
  {{Moncuquet}}, \citenamefont {{Monson}}, \citenamefont {{Mozer}},
  \citenamefont {{Murphy}}, \citenamefont {{Odom}}, \citenamefont
  {{Oliverson}}, \citenamefont {{Olson}}, \citenamefont {{Parker}},
  \citenamefont {{Pankow}}, \citenamefont {{Phan}}, \citenamefont {{Quataert}},
  \citenamefont {{Quinn}}, \citenamefont {{Ruplin}}, \citenamefont {{Salem}},
  \citenamefont {{Seitz}}, \citenamefont {{Sheppard}}, \citenamefont {{Siy}},
  \citenamefont {{Stevens}}, \citenamefont {{Summers}}, \citenamefont
  {{Szabo}}, \citenamefont {{Timofeeva}}, \citenamefont {{Vaivads}},
  \citenamefont {{Velli}}, \citenamefont {{Yehle}}, \citenamefont
  {{Werthimer}},\ and\ \citenamefont {{Wygant}}}]{Bale_al_2016}%
  \BibitemOpen
  \bibfield  {author} {\bibinfo {author} {\bibfnamefont {S.~D.}\ \bibnamefont
  {{Bale}}}, \bibinfo {author} {\bibfnamefont {K.}~\bibnamefont {{Goetz}}},
  \bibinfo {author} {\bibfnamefont {P.~R.}\ \bibnamefont {{Harvey}}}, \bibinfo
  {author} {\bibfnamefont {P.}~\bibnamefont {{Turin}}}, \bibinfo {author}
  {\bibfnamefont {J.~W.}\ \bibnamefont {{Bonnell}}}, \bibinfo {author}
  {\bibfnamefont {T.}~\bibnamefont {{Dudok de Wit}}}, \bibinfo {author}
  {\bibfnamefont {R.~E.}\ \bibnamefont {{Ergun}}}, \bibinfo {author}
  {\bibfnamefont {R.~J.}\ \bibnamefont {{MacDowall}}}, \bibinfo {author}
  {\bibfnamefont {M.}~\bibnamefont {{Pulupa}}}, \bibinfo {author}
  {\bibfnamefont {M.}~\bibnamefont {{Andre}}}, \bibinfo {author} {\bibfnamefont
  {M.}~\bibnamefont {{Bolton}}}, \bibinfo {author} {\bibfnamefont {J.~L.}\
  \bibnamefont {{Bougeret}}}, \bibinfo {author} {\bibfnamefont {T.~A.}\
  \bibnamefont {{Bowen}}}, \bibinfo {author} {\bibfnamefont {D.}~\bibnamefont
  {{Burgess}}}, \bibinfo {author} {\bibfnamefont {C.~A.}\ \bibnamefont
  {{Cattell}}}, \bibinfo {author} {\bibfnamefont {B.~D.~G.}\ \bibnamefont
  {{Chandran}}}, \bibinfo {author} {\bibfnamefont {C.~C.}\ \bibnamefont
  {{Chaston}}}, \bibinfo {author} {\bibfnamefont {C.~H.~K.}\ \bibnamefont
  {{Chen}}}, \bibinfo {author} {\bibfnamefont {M.~K.}\ \bibnamefont {{Choi}}},
  \bibinfo {author} {\bibfnamefont {J.~E.}\ \bibnamefont {{Connerney}}},
  \bibinfo {author} {\bibfnamefont {S.}~\bibnamefont {{Cranmer}}}, \bibinfo
  {author} {\bibfnamefont {M.}~\bibnamefont {{Diaz-Aguado}}}, \bibinfo {author}
  {\bibfnamefont {W.}~\bibnamefont {{Donakowski}}}, \bibinfo {author}
  {\bibfnamefont {J.~F.}\ \bibnamefont {{Drake}}}, \bibinfo {author}
  {\bibfnamefont {W.~M.}\ \bibnamefont {{Farrell}}}, \bibinfo {author}
  {\bibfnamefont {P.}~\bibnamefont {{Fergeau}}}, \bibinfo {author}
  {\bibfnamefont {J.}~\bibnamefont {{Fermin}}}, \bibinfo {author}
  {\bibfnamefont {J.}~\bibnamefont {{Fischer}}}, \bibinfo {author}
  {\bibfnamefont {N.}~\bibnamefont {{Fox}}}, \bibinfo {author} {\bibfnamefont
  {D.}~\bibnamefont {{Glaser}}}, \bibinfo {author} {\bibfnamefont
  {M.}~\bibnamefont {{Goldstein}}}, \bibinfo {author} {\bibfnamefont
  {D.}~\bibnamefont {{Gordon}}}, \bibinfo {author} {\bibfnamefont
  {E.}~\bibnamefont {{Hanson}}}, \bibinfo {author} {\bibfnamefont {S.~E.}\
  \bibnamefont {{Harris}}}, \bibinfo {author} {\bibfnamefont {L.~M.}\
  \bibnamefont {{Hayes}}}, \bibinfo {author} {\bibfnamefont {J.~J.}\
  \bibnamefont {{Hinze}}}, \bibinfo {author} {\bibfnamefont {J.~V.}\
  \bibnamefont {{Hollweg}}}, \bibinfo {author} {\bibfnamefont {T.~S.}\
  \bibnamefont {{Horbury}}}, \bibinfo {author} {\bibfnamefont {R.~A.}\
  \bibnamefont {{Howard}}}, \bibinfo {author} {\bibfnamefont {V.}~\bibnamefont
  {{Hoxie}}}, \bibinfo {author} {\bibfnamefont {G.}~\bibnamefont {{Jannet}}},
  \bibinfo {author} {\bibfnamefont {M.}~\bibnamefont {{Karlsson}}}, \bibinfo
  {author} {\bibfnamefont {J.~C.}\ \bibnamefont {{Kasper}}}, \bibinfo {author}
  {\bibfnamefont {P.~J.}\ \bibnamefont {{Kellogg}}}, \bibinfo {author}
  {\bibfnamefont {M.}~\bibnamefont {{Kien}}}, \bibinfo {author} {\bibfnamefont
  {J.~A.}\ \bibnamefont {{Klimchuk}}}, \bibinfo {author} {\bibfnamefont
  {V.~V.}\ \bibnamefont {{Krasnoselskikh}}}, \bibinfo {author} {\bibfnamefont
  {S.}~\bibnamefont {{Krucker}}}, \bibinfo {author} {\bibfnamefont {J.~J.}\
  \bibnamefont {{Lynch}}}, \bibinfo {author} {\bibfnamefont {M.}~\bibnamefont
  {{Maksimovic}}}, \bibinfo {author} {\bibfnamefont {D.~M.}\ \bibnamefont
  {{Malaspina}}}, \bibinfo {author} {\bibfnamefont {S.}~\bibnamefont
  {{Marker}}}, \bibinfo {author} {\bibfnamefont {P.}~\bibnamefont {{Martin}}},
  \bibinfo {author} {\bibfnamefont {J.}~\bibnamefont {{Martinez-Oliveros}}},
  \bibinfo {author} {\bibfnamefont {J.}~\bibnamefont {{McCauley}}}, \bibinfo
  {author} {\bibfnamefont {D.~J.}\ \bibnamefont {{McComas}}}, \bibinfo {author}
  {\bibfnamefont {T.}~\bibnamefont {{McDonald}}}, \bibinfo {author}
  {\bibfnamefont {N.}~\bibnamefont {{Meyer-Vernet}}}, \bibinfo {author}
  {\bibfnamefont {M.}~\bibnamefont {{Moncuquet}}}, \bibinfo {author}
  {\bibfnamefont {S.~J.}\ \bibnamefont {{Monson}}}, \bibinfo {author}
  {\bibfnamefont {F.~S.}\ \bibnamefont {{Mozer}}}, \bibinfo {author}
  {\bibfnamefont {S.~D.}\ \bibnamefont {{Murphy}}}, \bibinfo {author}
  {\bibfnamefont {J.}~\bibnamefont {{Odom}}}, \bibinfo {author} {\bibfnamefont
  {R.}~\bibnamefont {{Oliverson}}}, \bibinfo {author} {\bibfnamefont
  {J.}~\bibnamefont {{Olson}}}, \bibinfo {author} {\bibfnamefont {E.~N.}\
  \bibnamefont {{Parker}}}, \bibinfo {author} {\bibfnamefont {D.}~\bibnamefont
  {{Pankow}}}, \bibinfo {author} {\bibfnamefont {T.}~\bibnamefont {{Phan}}},
  \bibinfo {author} {\bibfnamefont {E.}~\bibnamefont {{Quataert}}}, \bibinfo
  {author} {\bibfnamefont {T.}~\bibnamefont {{Quinn}}}, \bibinfo {author}
  {\bibfnamefont {S.~W.}\ \bibnamefont {{Ruplin}}}, \bibinfo {author}
  {\bibfnamefont {C.}~\bibnamefont {{Salem}}}, \bibinfo {author} {\bibfnamefont
  {D.}~\bibnamefont {{Seitz}}}, \bibinfo {author} {\bibfnamefont {D.~A.}\
  \bibnamefont {{Sheppard}}}, \bibinfo {author} {\bibfnamefont
  {A.}~\bibnamefont {{Siy}}}, \bibinfo {author} {\bibfnamefont
  {K.}~\bibnamefont {{Stevens}}}, \bibinfo {author} {\bibfnamefont
  {D.}~\bibnamefont {{Summers}}}, \bibinfo {author} {\bibfnamefont
  {A.}~\bibnamefont {{Szabo}}}, \bibinfo {author} {\bibfnamefont
  {M.}~\bibnamefont {{Timofeeva}}}, \bibinfo {author} {\bibfnamefont
  {A.}~\bibnamefont {{Vaivads}}}, \bibinfo {author} {\bibfnamefont
  {M.}~\bibnamefont {{Velli}}}, \bibinfo {author} {\bibfnamefont
  {A.}~\bibnamefont {{Yehle}}}, \bibinfo {author} {\bibfnamefont
  {D.}~\bibnamefont {{Werthimer}}},\ and\ \bibinfo {author} {\bibfnamefont
  {J.~R.}\ \bibnamefont {{Wygant}}},\ }\bibfield  {title} {\bibinfo {title}
  {{The FIELDS Instrument Suite for Solar Probe Plus. Measuring the Coronal
  Plasma and Magnetic Field, Plasma Waves and Turbulence, and Radio Signatures
  of Solar Transients}},\ }\href {https://doi.org/10.1007/s11214-016-0244-5}
  {\bibfield  {journal} {\bibinfo  {journal} {Space Sci. Rev.}\ }\textbf
  {\bibinfo {volume} {204}},\ \bibinfo {pages} {49} (\bibinfo {year}
  {2016})}\BibitemShut {NoStop}%
\bibitem [{\citenamefont {Chen}\ \emph {et~al.}(2020)\citenamefont {Chen},
  \citenamefont {Bale}, \citenamefont {Bonnell}, \citenamefont {Borovikov},
  \citenamefont {Bowen}, \citenamefont {Burgess}, \citenamefont {Case},
  \citenamefont {Chandran}, \citenamefont {de~Wit}, \citenamefont {Goetz},
  \citenamefont {Harvey}, \citenamefont {Kasper}, \citenamefont {Klein},
  \citenamefont {Korreck}, \citenamefont {Larson}, \citenamefont {Livi},
  \citenamefont {MacDowall}, \citenamefont {Malaspina}, \citenamefont {Mallet},
  \citenamefont {McManus}, \citenamefont {Moncuquet}, \citenamefont {Pulupa},
  \citenamefont {Stevens},\ and\ \citenamefont {Whittlesey}}]{Chen_al_2020}%
  \BibitemOpen
  \bibfield  {author} {\bibinfo {author} {\bibfnamefont {C.~H.~K.}\
  \bibnamefont {Chen}}, \bibinfo {author} {\bibfnamefont {S.~D.}\ \bibnamefont
  {Bale}}, \bibinfo {author} {\bibfnamefont {J.~W.}\ \bibnamefont {Bonnell}},
  \bibinfo {author} {\bibfnamefont {D.}~\bibnamefont {Borovikov}}, \bibinfo
  {author} {\bibfnamefont {T.~A.}\ \bibnamefont {Bowen}}, \bibinfo {author}
  {\bibfnamefont {D.}~\bibnamefont {Burgess}}, \bibinfo {author} {\bibfnamefont
  {A.~W.}\ \bibnamefont {Case}}, \bibinfo {author} {\bibfnamefont {B.~D.~G.}\
  \bibnamefont {Chandran}}, \bibinfo {author} {\bibfnamefont {T.~D.}\
  \bibnamefont {de~Wit}}, \bibinfo {author} {\bibfnamefont {K.}~\bibnamefont
  {Goetz}}, \bibinfo {author} {\bibfnamefont {P.~R.}\ \bibnamefont {Harvey}},
  \bibinfo {author} {\bibfnamefont {J.~C.}\ \bibnamefont {Kasper}}, \bibinfo
  {author} {\bibfnamefont {K.~G.}\ \bibnamefont {Klein}}, \bibinfo {author}
  {\bibfnamefont {K.~E.}\ \bibnamefont {Korreck}}, \bibinfo {author}
  {\bibfnamefont {D.}~\bibnamefont {Larson}}, \bibinfo {author} {\bibfnamefont
  {R.}~\bibnamefont {Livi}}, \bibinfo {author} {\bibfnamefont {R.~J.}\
  \bibnamefont {MacDowall}}, \bibinfo {author} {\bibfnamefont {D.~M.}\
  \bibnamefont {Malaspina}}, \bibinfo {author} {\bibfnamefont {A.}~\bibnamefont
  {Mallet}}, \bibinfo {author} {\bibfnamefont {M.~D.}\ \bibnamefont {McManus}},
  \bibinfo {author} {\bibfnamefont {M.}~\bibnamefont {Moncuquet}}, \bibinfo
  {author} {\bibfnamefont {M.}~\bibnamefont {Pulupa}}, \bibinfo {author}
  {\bibfnamefont {M.~L.}\ \bibnamefont {Stevens}},\ and\ \bibinfo {author}
  {\bibfnamefont {P.}~\bibnamefont {Whittlesey}},\ }\bibfield  {title}
  {\bibinfo {title} {The evolution and role of solar wind turbulence in the
  inner heliosphere},\ }\href {https://doi.org/10.3847/1538-4365/ab60a3}
  {\bibfield  {journal} {\bibinfo  {journal} {Astrophys. J., Suppl. Ser.}\
  }\textbf {\bibinfo {volume} {246}},\ \bibinfo {pages} {53} (\bibinfo {year}
  {2020})}\BibitemShut {NoStop}%
\bibitem [{\citenamefont {Franci}\ \emph {et~al.}(2018)\citenamefont {Franci},
  \citenamefont {Hellinger}, \citenamefont {Guarrasi}, \citenamefont {Chen},
  \citenamefont {Papini}, \citenamefont {Verdini}, \citenamefont {Matteini},\
  and\ \citenamefont {Landi}}]{Franci_al_2018b}%
  \BibitemOpen
  \bibfield  {author} {\bibinfo {author} {\bibfnamefont {L.}~\bibnamefont
  {Franci}}, \bibinfo {author} {\bibfnamefont {P.}~\bibnamefont {Hellinger}},
  \bibinfo {author} {\bibfnamefont {M.}~\bibnamefont {Guarrasi}}, \bibinfo
  {author} {\bibfnamefont {C.~H.~K.}\ \bibnamefont {Chen}}, \bibinfo {author}
  {\bibfnamefont {E.}~\bibnamefont {Papini}}, \bibinfo {author} {\bibfnamefont
  {A.}~\bibnamefont {Verdini}}, \bibinfo {author} {\bibfnamefont
  {L.}~\bibnamefont {Matteini}},\ and\ \bibinfo {author} {\bibfnamefont
  {S.}~\bibnamefont {Landi}},\ }\bibfield  {title} {\bibinfo {title}
  {Three-dimensional simulations of solar wind turbulence with the hybrid code
  {CAMELIA}},\ }\href {https://doi.org/10.1088/1742-6596/1031/1/012002}
  {\bibfield  {journal} {\bibinfo  {journal} {J. Phys. Conf. Ser.}\ }\textbf
  {\bibinfo {volume} {1031}},\ \bibinfo {pages} {012002} (\bibinfo {year}
  {2018})}\BibitemShut {NoStop}%
\bibitem [{\citenamefont {Franci}\ \emph {et~al.}(2020)\citenamefont {Franci},
  \citenamefont {Stawarz}, \citenamefont {Papini}, \citenamefont {Hellinger},
  \citenamefont {Nakamura}, \citenamefont {Burgess}, \citenamefont {Landi},
  \citenamefont {Verdini}, \citenamefont {Matteini}, \citenamefont {Ergun},
  \citenamefont {Contel},\ and\ \citenamefont {Lindqvist}}]{Franci_al_2020}%
  \BibitemOpen
  \bibfield  {author} {\bibinfo {author} {\bibfnamefont {L.}~\bibnamefont
  {Franci}}, \bibinfo {author} {\bibfnamefont {J.~E.}\ \bibnamefont {Stawarz}},
  \bibinfo {author} {\bibfnamefont {E.}~\bibnamefont {Papini}}, \bibinfo
  {author} {\bibfnamefont {P.}~\bibnamefont {Hellinger}}, \bibinfo {author}
  {\bibfnamefont {T.}~\bibnamefont {Nakamura}}, \bibinfo {author}
  {\bibfnamefont {D.}~\bibnamefont {Burgess}}, \bibinfo {author} {\bibfnamefont
  {S.}~\bibnamefont {Landi}}, \bibinfo {author} {\bibfnamefont
  {A.}~\bibnamefont {Verdini}}, \bibinfo {author} {\bibfnamefont
  {L.}~\bibnamefont {Matteini}}, \bibinfo {author} {\bibfnamefont
  {R.}~\bibnamefont {Ergun}}, \bibinfo {author} {\bibfnamefont {O.~L.}\
  \bibnamefont {Contel}},\ and\ \bibinfo {author} {\bibfnamefont {P.-A.}\
  \bibnamefont {Lindqvist}},\ }\bibfield  {title} {\bibinfo {title} {Modeling
  {MMS} observations at the earth's magnetopause with hybrid simulations of
  alfv{\'{e}}nic turbulence},\ }\href
  {https://doi.org/10.3847/1538-4357/ab9a47} {\bibfield  {journal} {\bibinfo
  {journal} {Astrophys. J.}\ }\textbf {\bibinfo {volume} {898}},\ \bibinfo
  {pages} {175} (\bibinfo {year} {2020})}\BibitemShut {NoStop}%
\bibitem [{\citenamefont {{Franci}}\ \emph {et~al.}(2018)\citenamefont
  {{Franci}}, \citenamefont {{Landi}}, \citenamefont {{Verdini}}, \citenamefont
  {{Matteini}},\ and\ \citenamefont {{Hellinger}}}]{Franci_al_2018a}%
  \BibitemOpen
  \bibfield  {author} {\bibinfo {author} {\bibfnamefont {L.}~\bibnamefont
  {{Franci}}}, \bibinfo {author} {\bibfnamefont {S.}~\bibnamefont {{Landi}}},
  \bibinfo {author} {\bibfnamefont {A.}~\bibnamefont {{Verdini}}}, \bibinfo
  {author} {\bibfnamefont {L.}~\bibnamefont {{Matteini}}},\ and\ \bibinfo
  {author} {\bibfnamefont {P.}~\bibnamefont {{Hellinger}}},\ }\bibfield
  {title} {\bibinfo {title} {{Solar Wind Turbulent Cascade from MHD to Sub-ion
  Scales: Large-size 3D Hybrid Particle-in-cell Simulations}},\ }\href
  {https://doi.org/10.3847/1538-4357/aaa3e8} {\bibfield  {journal} {\bibinfo
  {journal} {Astrophys.~J.}\ }\textbf {\bibinfo {volume} {853}},\ \bibinfo
  {eid} {26} (\bibinfo {year} {2018})}\BibitemShut {NoStop}%
\bibitem [{\citenamefont {{{\v S}afr{\'a}nkov{\'a}}}\ \emph
  {et~al.}(2016)\citenamefont {{{\v S}afr{\'a}nkov{\'a}}}, \citenamefont {{N{\v
  e}me{\v c}ek}}, \citenamefont {{N{\v e}mec}}, \citenamefont {{P{\v r}ech}},
  \citenamefont {{Chen}},\ and\ \citenamefont
  {{Zastenker}}}]{Safrankova_al_2016}%
  \BibitemOpen
  \bibfield  {author} {\bibinfo {author} {\bibfnamefont {J.}~\bibnamefont {{{\v
  S}afr{\'a}nkov{\'a}}}}, \bibinfo {author} {\bibfnamefont {Z.}~\bibnamefont
  {{N{\v e}me{\v c}ek}}}, \bibinfo {author} {\bibfnamefont {F.}~\bibnamefont
  {{N{\v e}mec}}}, \bibinfo {author} {\bibfnamefont {L.}~\bibnamefont {{P{\v
  r}ech}}}, \bibinfo {author} {\bibfnamefont {C.~H.~K.}\ \bibnamefont
  {{Chen}}},\ and\ \bibinfo {author} {\bibfnamefont {G.~N.}\ \bibnamefont
  {{Zastenker}}},\ }\bibfield  {title} {\bibinfo {title} {{Power Spectral
  Density of Fluctuations of Bulk and Thermal Speeds in the Solar Wind}},\
  }\href {https://doi.org/10.3847/0004-637X/825/2/121} {\bibfield  {journal}
  {\bibinfo  {journal} {Astrophys.~J.}\ }\textbf {\bibinfo {volume} {825}},\
  \bibinfo {eid} {121} (\bibinfo {year} {2016})}\BibitemShut {NoStop}%
\bibitem [{\citenamefont {{Stawarz}}\ \emph {et~al.}(2016)\citenamefont
  {{Stawarz}}, \citenamefont {{Eriksson}}, \citenamefont {{Wilder}},
  \citenamefont {{Ergun}}, \citenamefont {{Schwartz}}, \citenamefont
  {{Pouquet}}, \citenamefont {{Burch}}, \citenamefont {{Giles}}, \citenamefont
  {{Khotyaintsev}}, \citenamefont {{Contel}}, \citenamefont {{Lindqvist}},
  \citenamefont {{Magnes}}, \citenamefont {{Pollock}}, \citenamefont
  {{Russell}}, \citenamefont {{Strangeway}}, \citenamefont {{Torbert}},
  \citenamefont {{Avanov}}, \citenamefont {{Dorelli}}, \citenamefont
  {{Eastwood}}, \citenamefont {{Gershman}}, \citenamefont {{Goodrich}},
  \citenamefont {{Malaspina}}, \citenamefont {{Marklund}}, \citenamefont
  {{Mirioni}},\ and\ \citenamefont {{Sturner}}}]{Stawarz_al_2016}%
  \BibitemOpen
  \bibfield  {author} {\bibinfo {author} {\bibfnamefont {J.~E.}\ \bibnamefont
  {{Stawarz}}}, \bibinfo {author} {\bibfnamefont {S.}~\bibnamefont
  {{Eriksson}}}, \bibinfo {author} {\bibfnamefont {F.~D.}\ \bibnamefont
  {{Wilder}}}, \bibinfo {author} {\bibfnamefont {R.~E.}\ \bibnamefont
  {{Ergun}}}, \bibinfo {author} {\bibfnamefont {S.~J.}\ \bibnamefont
  {{Schwartz}}}, \bibinfo {author} {\bibfnamefont {A.}~\bibnamefont
  {{Pouquet}}}, \bibinfo {author} {\bibfnamefont {J.~L.}\ \bibnamefont
  {{Burch}}}, \bibinfo {author} {\bibfnamefont {B.~L.}\ \bibnamefont
  {{Giles}}}, \bibinfo {author} {\bibfnamefont {Y.}~\bibnamefont
  {{Khotyaintsev}}}, \bibinfo {author} {\bibfnamefont {O.~L.}\ \bibnamefont
  {{Contel}}}, \bibinfo {author} {\bibfnamefont {P.-A.}\ \bibnamefont
  {{Lindqvist}}}, \bibinfo {author} {\bibfnamefont {W.}~\bibnamefont
  {{Magnes}}}, \bibinfo {author} {\bibfnamefont {C.~J.}\ \bibnamefont
  {{Pollock}}}, \bibinfo {author} {\bibfnamefont {C.~T.}\ \bibnamefont
  {{Russell}}}, \bibinfo {author} {\bibfnamefont {R.~J.}\ \bibnamefont
  {{Strangeway}}}, \bibinfo {author} {\bibfnamefont {R.~B.}\ \bibnamefont
  {{Torbert}}}, \bibinfo {author} {\bibfnamefont {L.~A.}\ \bibnamefont
  {{Avanov}}}, \bibinfo {author} {\bibfnamefont {J.~C.}\ \bibnamefont
  {{Dorelli}}}, \bibinfo {author} {\bibfnamefont {J.~P.}\ \bibnamefont
  {{Eastwood}}}, \bibinfo {author} {\bibfnamefont {D.~J.}\ \bibnamefont
  {{Gershman}}}, \bibinfo {author} {\bibfnamefont {K.~A.}\ \bibnamefont
  {{Goodrich}}}, \bibinfo {author} {\bibfnamefont {D.~M.}\ \bibnamefont
  {{Malaspina}}}, \bibinfo {author} {\bibfnamefont {G.~T.}\ \bibnamefont
  {{Marklund}}}, \bibinfo {author} {\bibfnamefont {L.}~\bibnamefont
  {{Mirioni}}},\ and\ \bibinfo {author} {\bibfnamefont {A.~P.}\ \bibnamefont
  {{Sturner}}},\ }\bibfield  {title} {\bibinfo {title} {{Observations of
  turbulence in a Kelvin-Helmholtz event on 8 September 2015 by the
  Magnetospheric Multiscale mission}},\ }\href
  {https://doi.org/10.1002/2016JA023458} {\bibfield  {journal} {\bibinfo
  {journal} {J.~Geophys.~Res.}\ }\textbf {\bibinfo {volume} {121}},\ \bibinfo
  {pages} {11} (\bibinfo {year} {2016})}\BibitemShut {NoStop}%
\bibitem [{\citenamefont {{Chen}}\ and\ \citenamefont
  {{Boldyrev}}(2017)}]{Chen_Boldyrev_2017}%
  \BibitemOpen
  \bibfield  {author} {\bibinfo {author} {\bibfnamefont {C.~H.~K.}\
  \bibnamefont {{Chen}}}\ and\ \bibinfo {author} {\bibfnamefont
  {S.}~\bibnamefont {{Boldyrev}}},\ }\bibfield  {title} {\bibinfo {title}
  {{Nature of Kinetic Scale Turbulence in the Earth's Magnetosheath}},\ }\href
  {https://doi.org/10.3847/1538-4357/aa74e0} {\bibfield  {journal} {\bibinfo
  {journal} {Astrophys.~J.}\ }\textbf {\bibinfo {volume} {842}},\ \bibinfo
  {eid} {122} (\bibinfo {year} {2017})}\BibitemShut {NoStop}%
\bibitem [{\citenamefont {Kingsep}\ \emph {et~al.}(1990)\citenamefont
  {Kingsep}, \citenamefont {Chukbar},\ and\ \citenamefont
  {{Yan'kov}}}]{Kingsep_al_1990}%
  \BibitemOpen
  \bibfield  {author} {\bibinfo {author} {\bibfnamefont {A.~S.}\ \bibnamefont
  {Kingsep}}, \bibinfo {author} {\bibfnamefont {K.~V.}\ \bibnamefont
  {Chukbar}},\ and\ \bibinfo {author} {\bibfnamefont {V.~V.}\ \bibnamefont
  {{Yan'kov}}},\ }\bibinfo {title} {Electron magnetohydrodynamics},\ in\
  \href@noop {} {\emph {\bibinfo {booktitle} {Reviews of Plasma Physics}}},\
  \bibinfo {editor} {edited by\ \bibinfo {editor} {\bibfnamefont {B.~B.}\
  \bibnamefont {Kadomtsev}}}\ (\bibinfo  {publisher} {Consultant Bureau},\
  \bibinfo {year} {1990})\ Chap.~\bibinfo {chapter} {16}, p.\ \bibinfo {pages}
  {243}\BibitemShut {NoStop}%
\bibitem [{\citenamefont {{Biskamp}}\ \emph {et~al.}(1996)\citenamefont
  {{Biskamp}}, \citenamefont {{Schwarz}},\ and\ \citenamefont
  {{Drake}}}]{Biskamp_al_1996}%
  \BibitemOpen
  \bibfield  {author} {\bibinfo {author} {\bibfnamefont {D.}~\bibnamefont
  {{Biskamp}}}, \bibinfo {author} {\bibfnamefont {E.}~\bibnamefont
  {{Schwarz}}},\ and\ \bibinfo {author} {\bibfnamefont {J.~F.}\ \bibnamefont
  {{Drake}}},\ }\bibfield  {title} {\bibinfo {title} {{Two-Dimensional Electron
  Magnetohydrodynamic Turbulence}},\ }\href
  {https://doi.org/10.1103/PhysRevLett.76.1264} {\bibfield  {journal} {\bibinfo
   {journal} {Phys. Rev. Lett.}\ }\textbf {\bibinfo {volume} {76}},\ \bibinfo
  {pages} {1264} (\bibinfo {year} {1996})}\BibitemShut {NoStop}%
\bibitem [{\citenamefont {Meyrand}\ and\ \citenamefont
  {Galtier}(2012)}]{Meyrand_Galtier_2012}%
  \BibitemOpen
  \bibfield  {author} {\bibinfo {author} {\bibfnamefont {R.}~\bibnamefont
  {Meyrand}}\ and\ \bibinfo {author} {\bibfnamefont {S.}~\bibnamefont
  {Galtier}},\ }\bibfield  {title} {\bibinfo {title} {Spontaneous chiral
  symmetry breaking of hall magnetohydrodynamic turbulence},\ }\href
  {https://doi.org/10.1103/PhysRevLett.109.194501} {\bibfield  {journal}
  {\bibinfo  {journal} {Phys. Rev. Lett.}\ }\textbf {\bibinfo {volume} {109}},\
  \bibinfo {pages} {194501} (\bibinfo {year} {2012})}\BibitemShut {NoStop}%
\bibitem [{\citenamefont {Franci}\ and\ \citenamefont
  {Sarto}(2020)}]{Franci_DelSarto_2020}%
  \BibitemOpen
  \bibfield  {author} {\bibinfo {author} {\bibfnamefont {L.}~\bibnamefont
  {Franci}}\ and\ \bibinfo {author} {\bibfnamefont {D.~D.}\ \bibnamefont
  {Sarto}},\ }\bibfield  {title} {\bibinfo {title} {Multiple "fluid ion-scale"
  regimes of self-similar energy transfer in plasma turbulence}} (\bibinfo
  {year} {2020}),\ \bibinfo {note} {to be submitted}\BibitemShut {NoStop}%
\bibitem [{\citenamefont {Galtier}\ and\ \citenamefont
  {Buchlin}(2007)}]{Galtier_Buchlin_2007}%
  \BibitemOpen
  \bibfield  {author} {\bibinfo {author} {\bibfnamefont {S.}~\bibnamefont
  {Galtier}}\ and\ \bibinfo {author} {\bibfnamefont {E.}~\bibnamefont
  {Buchlin}},\ }\bibfield  {title} {\bibinfo {title} {Multiscale
  hall-magnetohydrodynamic turbulence in the solar wind},\ }\href
  {https://doi.org/10.1086/510423} {\bibfield  {journal} {\bibinfo  {journal}
  {Astrophys. J.}\ }\textbf {\bibinfo {volume} {656}},\ \bibinfo {pages} {560}
  (\bibinfo {year} {2007})}\BibitemShut {NoStop}%
\bibitem [{\citenamefont {{Hellinger}}\ \emph {et~al.}(2018)\citenamefont
  {{Hellinger}}, \citenamefont {{Verdini}}, \citenamefont {{Landi}},
  \citenamefont {{Franci}},\ and\ \citenamefont
  {{Matteini}}}]{Hellinger_al_2018}%
  \BibitemOpen
  \bibfield  {author} {\bibinfo {author} {\bibfnamefont {P.}~\bibnamefont
  {{Hellinger}}}, \bibinfo {author} {\bibfnamefont {A.}~\bibnamefont
  {{Verdini}}}, \bibinfo {author} {\bibfnamefont {S.}~\bibnamefont {{Landi}}},
  \bibinfo {author} {\bibfnamefont {L.}~\bibnamefont {{Franci}}},\ and\
  \bibinfo {author} {\bibfnamefont {L.}~\bibnamefont {{Matteini}}},\ }\bibfield
   {title} {\bibinfo {title} {{von K{\'a}rm{\'a}n-Howarth Equation for Hall
  Magnetohydrodynamics: Hybrid Simulations}},\ }\href
  {https://doi.org/10.3847/2041-8213/aabc06} {\bibfield  {journal} {\bibinfo
  {journal} {Astrophys.~J.~Lett.}\ }\textbf {\bibinfo {volume} {857}},\
  \bibinfo {eid} {L19} (\bibinfo {year} {2018})}\BibitemShut {NoStop}%
\bibitem [{\citenamefont {{Franci}}\ \emph {et~al.}(2016)\citenamefont
  {{Franci}}, \citenamefont {{Landi}}, \citenamefont {{Matteini}},
  \citenamefont {{Verdini}},\ and\ \citenamefont
  {{Hellinger}}}]{Franci_al_2016b}%
  \BibitemOpen
  \bibfield  {author} {\bibinfo {author} {\bibfnamefont {L.}~\bibnamefont
  {{Franci}}}, \bibinfo {author} {\bibfnamefont {S.}~\bibnamefont {{Landi}}},
  \bibinfo {author} {\bibfnamefont {L.}~\bibnamefont {{Matteini}}}, \bibinfo
  {author} {\bibfnamefont {A.}~\bibnamefont {{Verdini}}},\ and\ \bibinfo
  {author} {\bibfnamefont {P.}~\bibnamefont {{Hellinger}}},\ }\bibfield
  {title} {\bibinfo {title} {{Plasma Beta Dependence of the Ion-scale Spectral
  Break of Solar Wind Turbulence: High-resolution 2D Hybrid Simulations}},\
  }\href {https://doi.org/10.3847/1538-4357/833/1/91} {\bibfield  {journal}
  {\bibinfo  {journal} {Astrophys.~J.}\ }\textbf {\bibinfo {volume} {833}},\
  \bibinfo {eid} {91} (\bibinfo {year} {2016})}\BibitemShut {NoStop}%
\bibitem [{\citenamefont {{Bruno}}\ \emph {et~al.}(2014)\citenamefont
  {{Bruno}}, \citenamefont {{Trenchi}},\ and\ \citenamefont
  {{Telloni}}}]{Bruno_al_2014}%
  \BibitemOpen
  \bibfield  {author} {\bibinfo {author} {\bibfnamefont {R.}~\bibnamefont
  {{Bruno}}}, \bibinfo {author} {\bibfnamefont {L.}~\bibnamefont {{Trenchi}}},\
  and\ \bibinfo {author} {\bibfnamefont {D.}~\bibnamefont {{Telloni}}},\
  }\bibfield  {title} {\bibinfo {title} {{Spectral Slope Variation at Proton
  Scales from Fast to Slow Solar Wind}},\ }\href
  {https://doi.org/10.1088/2041-8205/793/1/L15} {\bibfield  {journal} {\bibinfo
   {journal} {Astrophys.~J.~Lett.}\ }\textbf {\bibinfo {volume} {793}},\
  \bibinfo {eid} {L15} (\bibinfo {year} {2014})}\BibitemShut {NoStop}%
\bibitem [{\citenamefont {Beskin}\ and\ \citenamefont
  {Nokhrina}(2009)}]{Beskin_Nokhrina_2009}%
  \BibitemOpen
  \bibfield  {author} {\bibinfo {author} {\bibfnamefont {V.~S.}\ \bibnamefont
  {Beskin}}\ and\ \bibinfo {author} {\bibfnamefont {E.~E.}\ \bibnamefont
  {Nokhrina}},\ }\bibfield  {title} {\bibinfo {title} {{On the central core in
  MHD winds and jets}},\ }\href
  {https://doi.org/10.1111/j.1365-2966.2009.14964.x} {\bibfield  {journal}
  {\bibinfo  {journal} {Mon.~Not.~Roy.~Astron.~Soc.}\ }\textbf {\bibinfo
  {volume} {397}},\ \bibinfo {pages} {1486} (\bibinfo {year}
  {2009})}\BibitemShut {NoStop}%
\bibitem [{\citenamefont {{Machida}}\ \emph {et~al.}(2000)\citenamefont
  {{Machida}}, \citenamefont {{Hayashi}},\ and\ \citenamefont
  {{Matsumoto}}}]{Machida_al_2000}%
  \BibitemOpen
  \bibfield  {author} {\bibinfo {author} {\bibfnamefont {M.}~\bibnamefont
  {{Machida}}}, \bibinfo {author} {\bibfnamefont {M.~R.}\ \bibnamefont
  {{Hayashi}}},\ and\ \bibinfo {author} {\bibfnamefont {R.}~\bibnamefont
  {{Matsumoto}}},\ }\bibfield  {title} {\bibinfo {title} {{Global Simulations
  of Differentially Rotating Magnetized Disks: Formation of
  Low-{\ensuremath{\beta}} Filaments and Structured Coronae}},\ }\href
  {https://doi.org/10.1086/312553} {\bibfield  {journal} {\bibinfo  {journal}
  {Astrophys.~J.~Lett.}\ }\textbf {\bibinfo {volume} {532}},\ \bibinfo {pages}
  {L67} (\bibinfo {year} {2000})}\BibitemShut {NoStop}%
\bibitem [{\citenamefont {Zhang}\ \emph {et~al.}(2017)\citenamefont {Zhang},
  \citenamefont {Ma},\ and\ \citenamefont {Wang}}]{Zhang_al_2017}%
  \BibitemOpen
  \bibfield  {author} {\bibinfo {author} {\bibfnamefont {W.}~\bibnamefont
  {Zhang}}, \bibinfo {author} {\bibfnamefont {Z.~W.}\ \bibnamefont {Ma}},\ and\
  \bibinfo {author} {\bibfnamefont {S.}~\bibnamefont {Wang}},\ }\bibfield
  {title} {\bibinfo {title} {Hall effect on tearing mode instabilities in
  tokamak},\ }\href {https://doi.org/10.1063/1.5004430} {\bibfield  {journal}
  {\bibinfo  {journal} {Phys. Plasmas}\ }\textbf {\bibinfo {volume} {24}},\
  \bibinfo {pages} {102510} (\bibinfo {year} {2017})}\BibitemShut {NoStop}%
\bibitem [{\citenamefont {Meinecke}\ \emph {et~al.}(2015)\citenamefont
  {Meinecke}, \citenamefont {Tzeferacos}, \citenamefont {Bell}, \citenamefont
  {Bingham}, \citenamefont {Clarke}, \citenamefont {Churazov}, \citenamefont
  {Crowston}, \citenamefont {Doyle}, \citenamefont {Drake}, \citenamefont
  {Heathcote}, \citenamefont {Koenig}, \citenamefont {Kuramitsu}, \citenamefont
  {Kuranz}, \citenamefont {Lee}, \citenamefont {MacDonald}, \citenamefont
  {Murphy}, \citenamefont {Notley}, \citenamefont {Park}, \citenamefont
  {Pelka}, \citenamefont {Ravasio}, \citenamefont {Reville}, \citenamefont
  {Sakawa}, \citenamefont {Wan}, \citenamefont {Woolsey}, \citenamefont
  {Yurchak}, \citenamefont {Miniati}, \citenamefont {Schekochihin},
  \citenamefont {Lamb},\ and\ \citenamefont {Gregori}}]{Meinecke_al_2015}%
  \BibitemOpen
  \bibfield  {author} {\bibinfo {author} {\bibfnamefont {J.}~\bibnamefont
  {Meinecke}}, \bibinfo {author} {\bibfnamefont {P.}~\bibnamefont
  {Tzeferacos}}, \bibinfo {author} {\bibfnamefont {A.}~\bibnamefont {Bell}},
  \bibinfo {author} {\bibfnamefont {R.}~\bibnamefont {Bingham}}, \bibinfo
  {author} {\bibfnamefont {R.}~\bibnamefont {Clarke}}, \bibinfo {author}
  {\bibfnamefont {E.}~\bibnamefont {Churazov}}, \bibinfo {author}
  {\bibfnamefont {R.}~\bibnamefont {Crowston}}, \bibinfo {author}
  {\bibfnamefont {H.}~\bibnamefont {Doyle}}, \bibinfo {author} {\bibfnamefont
  {R.~P.}\ \bibnamefont {Drake}}, \bibinfo {author} {\bibfnamefont
  {R.}~\bibnamefont {Heathcote}}, \bibinfo {author} {\bibfnamefont
  {M.}~\bibnamefont {Koenig}}, \bibinfo {author} {\bibfnamefont
  {Y.}~\bibnamefont {Kuramitsu}}, \bibinfo {author} {\bibfnamefont
  {C.}~\bibnamefont {Kuranz}}, \bibinfo {author} {\bibfnamefont
  {D.}~\bibnamefont {Lee}}, \bibinfo {author} {\bibfnamefont {M.}~\bibnamefont
  {MacDonald}}, \bibinfo {author} {\bibfnamefont {C.}~\bibnamefont {Murphy}},
  \bibinfo {author} {\bibfnamefont {M.}~\bibnamefont {Notley}}, \bibinfo
  {author} {\bibfnamefont {H.-S.}\ \bibnamefont {Park}}, \bibinfo {author}
  {\bibfnamefont {A.}~\bibnamefont {Pelka}}, \bibinfo {author} {\bibfnamefont
  {A.}~\bibnamefont {Ravasio}}, \bibinfo {author} {\bibfnamefont
  {B.}~\bibnamefont {Reville}}, \bibinfo {author} {\bibfnamefont
  {Y.}~\bibnamefont {Sakawa}}, \bibinfo {author} {\bibfnamefont
  {W.}~\bibnamefont {Wan}}, \bibinfo {author} {\bibfnamefont {N.}~\bibnamefont
  {Woolsey}}, \bibinfo {author} {\bibfnamefont {R.}~\bibnamefont {Yurchak}},
  \bibinfo {author} {\bibfnamefont {F.}~\bibnamefont {Miniati}}, \bibinfo
  {author} {\bibfnamefont {A.}~\bibnamefont {Schekochihin}}, \bibinfo {author}
  {\bibfnamefont {D.}~\bibnamefont {Lamb}},\ and\ \bibinfo {author}
  {\bibfnamefont {G.}~\bibnamefont {Gregori}},\ }\bibfield  {title} {\bibinfo
  {title} {Developed turbulence and nonlinear amplification of magnetic fields
  in laboratory and astrophysical plasmas},\ }\href
  {https://doi.org/10.1073/pnas.1502079112} {\bibfield  {journal} {\bibinfo
  {journal} {Proc. Natl. Acad. Sci. USA}\ }\textbf {\bibinfo {volume} {112}},\
  \bibinfo {pages} {8211} (\bibinfo {year} {2015})}\BibitemShut {NoStop}%
\bibitem [{\citenamefont {Tzeferacos}\ \emph {et~al.}(2018)\citenamefont
  {Tzeferacos}, \citenamefont {Rigby}, \citenamefont {Bott}, \citenamefont
  {Bell}, \citenamefont {Bingham}, \citenamefont {Casner}, \citenamefont
  {Cattaneo}, \citenamefont {Churazov}, \citenamefont {Emig}, \citenamefont
  {Fiuza},\ and\ \citenamefont {et~al.}}]{Tzeferacos_al_2018}%
  \BibitemOpen
  \bibfield  {author} {\bibinfo {author} {\bibfnamefont {P.}~\bibnamefont
  {Tzeferacos}}, \bibinfo {author} {\bibfnamefont {A.}~\bibnamefont {Rigby}},
  \bibinfo {author} {\bibfnamefont {A.~F.~A.}\ \bibnamefont {Bott}}, \bibinfo
  {author} {\bibfnamefont {A.~R.}\ \bibnamefont {Bell}}, \bibinfo {author}
  {\bibfnamefont {R.}~\bibnamefont {Bingham}}, \bibinfo {author} {\bibfnamefont
  {A.}~\bibnamefont {Casner}}, \bibinfo {author} {\bibfnamefont
  {F.}~\bibnamefont {Cattaneo}}, \bibinfo {author} {\bibfnamefont {E.~M.}\
  \bibnamefont {Churazov}}, \bibinfo {author} {\bibfnamefont {J.}~\bibnamefont
  {Emig}}, \bibinfo {author} {\bibfnamefont {F.}~\bibnamefont {Fiuza}},\ and\
  \bibinfo {author} {\bibnamefont {et~al.}},\ }\bibfield  {title} {\bibinfo
  {title} {Laboratory evidence of dynamo amplification of magnetic fields in a
  turbulent plasma},\ }\href@noop {} {\bibfield  {journal} {\bibinfo  {journal}
  {Nat. Commun.}\ }\textbf {\bibinfo {volume} {9}} (\bibinfo {year}
  {2018})}\BibitemShut {NoStop}%
\bibitem [{\citenamefont {Matthews}(1994)}]{Matthews_1994}%
  \BibitemOpen
  \bibfield  {author} {\bibinfo {author} {\bibfnamefont {A.~P.}\ \bibnamefont
  {Matthews}},\ }\bibfield  {title} {\bibinfo {title} {Current advance method
  and cyclic leapfrog for 2d multispecies hybrid plasma simulations},\ }\href
  {https://doi.org/10.1006/jcph.1994.1084} {\bibfield  {journal} {\bibinfo
  {journal} {J. Comput. Phys.}\ }\textbf {\bibinfo {volume} {112}},\ \bibinfo
  {pages} {102 } (\bibinfo {year} {1994})}\BibitemShut {NoStop}%
\bibitem [{\citenamefont {Halekas}\ \emph {et~al.}(2020)\citenamefont
  {Halekas}, \citenamefont {Whittlesey}, \citenamefont {Larson}, \citenamefont
  {McGinnis}, \citenamefont {Maksimovic}, \citenamefont {Berthomier},
  \citenamefont {Kasper}, \citenamefont {Case}, \citenamefont {Korreck},
  \citenamefont {Stevens}, \citenamefont {Klein}, \citenamefont {Bale},
  \citenamefont {MacDowall}, \citenamefont {Pulupa}, \citenamefont {Malaspina},
  \citenamefont {Goetz},\ and\ \citenamefont {Harvey}}]{Halekas_al_2020}%
  \BibitemOpen
  \bibfield  {author} {\bibinfo {author} {\bibfnamefont {J.~S.}\ \bibnamefont
  {Halekas}}, \bibinfo {author} {\bibfnamefont {P.}~\bibnamefont {Whittlesey}},
  \bibinfo {author} {\bibfnamefont {D.~E.}\ \bibnamefont {Larson}}, \bibinfo
  {author} {\bibfnamefont {D.}~\bibnamefont {McGinnis}}, \bibinfo {author}
  {\bibfnamefont {M.}~\bibnamefont {Maksimovic}}, \bibinfo {author}
  {\bibfnamefont {M.}~\bibnamefont {Berthomier}}, \bibinfo {author}
  {\bibfnamefont {J.~C.}\ \bibnamefont {Kasper}}, \bibinfo {author}
  {\bibfnamefont {A.~W.}\ \bibnamefont {Case}}, \bibinfo {author}
  {\bibfnamefont {K.~E.}\ \bibnamefont {Korreck}}, \bibinfo {author}
  {\bibfnamefont {M.~L.}\ \bibnamefont {Stevens}}, \bibinfo {author}
  {\bibfnamefont {K.~G.}\ \bibnamefont {Klein}}, \bibinfo {author}
  {\bibfnamefont {S.~D.}\ \bibnamefont {Bale}}, \bibinfo {author}
  {\bibfnamefont {R.~J.}\ \bibnamefont {MacDowall}}, \bibinfo {author}
  {\bibfnamefont {M.~P.}\ \bibnamefont {Pulupa}}, \bibinfo {author}
  {\bibfnamefont {D.~M.}\ \bibnamefont {Malaspina}}, \bibinfo {author}
  {\bibfnamefont {K.}~\bibnamefont {Goetz}},\ and\ \bibinfo {author}
  {\bibfnamefont {P.~R.}\ \bibnamefont {Harvey}},\ }\bibfield  {title}
  {\bibinfo {title} {Electrons in the young solar wind: First results from the
  parker solar probe},\ }\href {https://doi.org/10.3847/1538-4365/ab4cec}
  {\bibfield  {journal} {\bibinfo  {journal} {Astrophys. J., Suppl. Ser.}\
  }\textbf {\bibinfo {volume} {246}},\ \bibinfo {pages} {22} (\bibinfo {year}
  {2020})}\BibitemShut {NoStop}%
\bibitem [{\citenamefont {{Franci}}\ \emph
  {et~al.}(2015{\natexlab{a}})\citenamefont {{Franci}}, \citenamefont
  {{Verdini}}, \citenamefont {{Matteini}}, \citenamefont {{Landi}},\ and\
  \citenamefont {{Hellinger}}}]{Franci_al_2015a}%
  \BibitemOpen
  \bibfield  {author} {\bibinfo {author} {\bibfnamefont {L.}~\bibnamefont
  {{Franci}}}, \bibinfo {author} {\bibfnamefont {A.}~\bibnamefont {{Verdini}}},
  \bibinfo {author} {\bibfnamefont {L.}~\bibnamefont {{Matteini}}}, \bibinfo
  {author} {\bibfnamefont {S.}~\bibnamefont {{Landi}}},\ and\ \bibinfo {author}
  {\bibfnamefont {P.}~\bibnamefont {{Hellinger}}},\ }\bibfield  {title}
  {\bibinfo {title} {Solar wind turbulence from mhd to sub-ion scales:
  high-resolution hybrid simulations},\ }\href@noop {} {\bibfield  {journal}
  {\bibinfo  {journal} {Astrophys.~J.~Lett.}\ }\textbf {\bibinfo {volume}
  {804}},\ \bibinfo {eid} {L39} (\bibinfo {year}
  {2015}{\natexlab{a}})}\BibitemShut {NoStop}%
\bibitem [{\citenamefont {{Franci}}\ \emph
  {et~al.}(2015{\natexlab{b}})\citenamefont {{Franci}}, \citenamefont
  {{Landi}}, \citenamefont {{Matteini}}, \citenamefont {{Verdini}},\ and\
  \citenamefont {{Hellinger}}}]{Franci_al_2015b}%
  \BibitemOpen
  \bibfield  {author} {\bibinfo {author} {\bibfnamefont {L.}~\bibnamefont
  {{Franci}}}, \bibinfo {author} {\bibfnamefont {S.}~\bibnamefont {{Landi}}},
  \bibinfo {author} {\bibfnamefont {L.}~\bibnamefont {{Matteini}}}, \bibinfo
  {author} {\bibfnamefont {A.}~\bibnamefont {{Verdini}}},\ and\ \bibinfo
  {author} {\bibfnamefont {P.}~\bibnamefont {{Hellinger}}},\ }\bibfield
  {title} {\bibinfo {title} {{High-resolution Hybrid Simulations of Kinetic
  Plasma Turbulence at Proton Scales}},\ }\href
  {https://doi.org/10.1088/0004-637X/812/1/21} {\bibfield  {journal} {\bibinfo
  {journal} {Astrophys.~J.}\ }\textbf {\bibinfo {volume} {812}},\ \bibinfo
  {eid} {21} (\bibinfo {year} {2015}{\natexlab{b}})}\BibitemShut {NoStop}%
\end{thebibliography}%

\section{Acknowledgements}
The authors acknowledge valuable discussions with Christopher Chen, Lorenzo Matteini, Andrea Verdini, and with the members of the FIELDS instrument team and of the Italian SWA/Solar Orbiter working groups.
L.F. and D.B. are supported by the UK Science and
Technology Facilities Council (STFC) grants ST/P000622/1
and ST/T00018X/1. L.F., D.D.S., E.P., and S.L. are grateful to Campus France and to the Universit\`a Italo-Francese for mobility funding provided through the bilateral projects PHC Galilee 2018-2019 N.39632WH   and Galileo G18\_569. J.E.S. is supported by the STFC grant
ST/S000364/1 and funded by the Royal Society University Research Fellowship URF{\textbackslash}R1{\textbackslash}201286. P.H. acknowledges grant 18-08861S of the
Czech Science Foundation. This work was performed using the Cambridge Service for Data Driven Discovery (CSD3), part of which is operated by the University of Cambridge Research Computing, and the DiRAC Data Intensive service at Leicester (DIaL), operated by the University of Leicester IT Services, both of which form part of the STFC DiRAC HPC Facility (www.dirac.ac.uk). The DiRAC component of CSD3 was funded by BEIS capital funding via STFC capital grants ST/P002307/1 and ST/R002452/1 and STFC operations grant ST/R00689X/1. The DIaL equipment was funded by BEIS capital funding via STFC capital grants ST/K000373/1 and ST/R002363/1 and STFC DiRAC Operations grant ST/R001014/1. DiRAC is part of the National e-Infrastructure.  Access to DiRAC resources was granted through Director's Discretionary Time allocations in 2019 and 2020. This research used resources of the Argonne Leadership Computing Facility, which is a DOE Office of Science User Facility supported under Contract DE-AC02-06CH11357 (Director's Discretionary project ``TurbLowBeta''). We acknowledge CINECA for the availability of high performance computing
resources and support under the program Accordo
Quadro INAF-CINECA 2017–2019 (grant C4A26). 
L.F. thanks Dr. Mark Wilkinson, director of the DiRAC HPC facility, for the support and the invitation to present the preliminary results of this work at the DiRAC Day 2019 and Christopher James Callow for the fundamental support during the lockdown due to the COVID-19 pandemic.

\section{Author contributions}
L.F. carried out the numerical simulations, performed the analysis of the numerical dataset, proposed the theoretical interpretation, produced the figures, and wrote the manuscript. D.D.S. contributed to the formulation of the theoretical framework and to the presentation of the results. E.P. performed the analysis of the observational dataset. A.G. selected the spacecraft data of interest and performed a preliminary analysis of the observational dataset. J.E. Stawarz suggested additional analysis in support of the physical interpretation of the results. All of the authors discussed the results and commented on the paper.

\section{Additional information}
\textbf{Competing interests:} The authors declare no competing financial interests.

\end{document}